\documentclass[%
 reprint,
 nofootinbib,
 amsmath,amssymb,
 aps,
 prd,
 superscriptaddress
]{revtex4-1}

\usepackage{graphicx}
\usepackage{dcolumn}
\usepackage{bm}

\usepackage{aas_macros}
\usepackage{color}
\usepackage{amsmath}
\usepackage{natbib}

\newcommand{\Msun}{\ensuremath{M_{\odot}}}

\begin{document}

\preprint{APS/123-QED}

\title{%
Detectability of Collective Neutrino Oscillation Signatures
\\ in the Supernova Explosion of a 8.8\Msun star}

\author{Hirokazu Sasaki}
\affiliation{%
Department of Astronomy Graduate School of Science The University of Tokyo, 7-3-1 Hongo, Bunkyo-ku, Tokyo 113-033, Japan}
\affiliation{%
Division of Science, National Astronomical Observatory of Japan, 2-21-1 Osawa, Mitaka, Tokyo 181-8588, Japan}%
\author{Tomoya Takiwaki}%
 \email{takiwaki.tomoya.astro@gmail.com, orcid: 0000-0003-0304-9283}
\affiliation{%
Division of Science, National Astronomical Observatory of Japan, 2-21-1 Osawa, Mitaka, Tokyo 181-8588, Japan}%
\author{Shio Kawagoe}
\affiliation{%
Institute of Industrial Science, The University of Tokyo, 4-6-1 Komaba, Meguro-ku, Tokyo 153-8505 Japan
}%
\author{Shunsaku Horiuchi}
 \email{horiuchi@vt.edu, orcid: 0000-0001-6142-6556} 
4

\affiliation{%
Center for Neutrino Physics, Department of Physics, Virginia Tech, Blacksburg, VA
24061-0435, United States of America
}%
\author{Koji Ishidoshiro}
\affiliation{%
Research Center for Neutrino Science, Tohoku University, Sendai 980-8578, Japan
}%

\date{\today}
\begin{abstract}
In order to investigate the impact of collective neutrino oscillations (CNO) on the neutrino signal from a nearby supernova, we perform 3-flavor neutrino oscillation simulations employing the multiangle effect. The background hydrodynamic model is based on the neutrino hydrodynamic simulation of a 8.8 \Msun progenitor star. We find that CNO commences after some 100 ms post bounce. Before this, CNO is suppressed by matter-induced decoherence. 
In the inverted mass hierarchy, the spectrum of $\bar{\nu}_e$ becomes softer after the onset of CNO. To evaluate the detectability of this modification, we define a hardness ratio between the number of high energy neutrino events and low energy neutrino events adopting a fixed critical energy. We show that Hyper-Kamiokande (HK) can distinguish the effect of CNO for supernova distances out to $\sim 10$ kpc. On the other hand, for the normal mass hierarchy, the spectrum of $\nu_e$ becomes softer after the onset of CNO, and we show that DUNE can distinguish this feature for supernova distances out to $\sim 10$ kpc. 
More work is necessary to optimize the best value of critical energy for maximum sensitivity.
We also show that if the spectrum of $\bar{\nu}_e$ in HK becomes softer due to CNO, the spectrum of $\nu_e$ in DUNE becomes harder, and vice versa. This synergistic observations in $\bar{\nu}_e$ and $\nu_e$, by HK and DUNE respectively, will be an intriguing opportunity to test the occurrence of CNO.
\end{abstract}

\maketitle


\section{Introduction}
A major goal of low-energy neutrino astronomy is observing 
a high statistics neutrino signal from a nearby core-collapse supernova \cite{Raffelt2012}.
The neutrino burst from SN1987A was an epoch-making observation that  
demonstrated the crucial link that massive stars
release huge amounts of energy at its endpoint in form of neutrinos and trigger a supernova explosion.
Even with limited neutrino event statistics (about 20 events), the signal revealed much about the importance of weak-interaction physics in the core-collapse.
Tens of thousands of neutrino events are expected from the next nearby core-collapse supernova, and the events are anticipated to provide unprecedented information of the explosion (see, e.g., the reviews \cite{Horiuchi2018WhatDetection,Janka2017,Hix2016TheSupernovae,Muller2016TheModels,Mirizzi2015SupernovaDetection,Foglizzo2015TheExperiments,Burrows2013ColloquiumTheory,Kotake2012Core-collapseRelativity}). 

There are multiple operational neutrino observatories that can detect a high-statistics neutrino burst event from a nearby supernova, e.g., 
SuperKamiokande (SK) \cite{Collaboration2007SearchSuper-Kamiokande}, IceCube \cite{Abbasi2011IceCubeSupernovae,Bruijn2013SupernovaFuture,Salathe2012NovelIceCube} and KamLAND \cite{Asakura2016KamLANDSTARS}.
SK is able to obtain both the light curve and spectrum of the neutrino. 
To decrease the background using coincident tagging technique,
SK will soon be upgraded with gadolinium \cite{Xu2016CurrentEGADS}. 
IceCube plays an important role in detecting time-variability of the signal with high statistics, by
taking advantage of its large volume \cite{Lund2010FastDetectability}.
KamLAND is sensitive to low-energy neutrino and is even sensitive to neutrinos emitted during the last
Si-burning phase of the progenitor \cite{Asakura2016KamLANDSTARS}.
In the future, large volume detectors such as 
HyperKamiokande (HK) \cite{Proto-Collaboration2018Hyper-KamiokandeReport} and JUNO \cite{Li2014OverviewJUNO} will become available for detecting electron anti-neutrinos out to further distances, while DUNE \cite{Acciarri2016} will dramatically enhance the capabilities to detect electron neutrinos.

In parallel to the development of neutrino observation facilities, the theory of neutrino emission and their propagation through the supernova and progenitor have dramatically progressed 
\cite{Horiuchi2018WhatDetection,Janka2017,Mirizzi2015SupernovaDetection}.
The technique of neutrino radiation hydrodynamics has become highly sophisticated \cite{Bruenn1985StellarEpoch,Liebendorfer2001ConservativeCoordinates,Rampp2002RadiationNeutrinos,Buras2006Two-dimensionalTransportb,Shibata2011TruncatedRelativity,Kuroda2016ASTARS,Sumiyoshi2012NEUTRINOCONFIGURATIONS} and can provide reliable neutrino luminosities and average energies.
Using reliable techniques,
three dimensional simulations are now available \cite{Takiwaki2012Three-dimensionalTransport,Takiwaki2014ASUPERNOVAE,Takiwaki2016Three-dimensionalFlows,Hanke2013SASICORES,Tamborra2014NeutrinoSimulations,Melson2015NEUTRINO-DRIVENSCATTERING,Melson2015Neutrino-drivenConvection,Vartanyan2019AModel} and
many interesting phenomena have been discovered, e.g., SASI \cite{Tamborra2014NeutrinoSimulations,Walk2018IdentifyingGyroscope}, LESA \cite{Tamborra2014SELF-SUSTAINEDDIMENSIONS}, and low-$T/|W|$ instability \cite{Takiwaki2018AnisotropicSupernovae}.

Perhaps the most distinct progress in the theory of neutrino propagation is the realization that  
Collective Neutrino Oscillations (CNO) can operate in the supernova environment (see, e.g., reviews \cite{Duan2009NeutrinoSupernovae} and references there in). 
CNO is an oscillation that happens in the high density region of neutrinos and leads to a potential for the flavor oscillation.
In this sense, CNO is similar to the MSW effect \cite{Wolfenstein1978NeutrinoMatter,Mikheev:1986gs}; however, the effect of CNO is complicated since the potential is made by the neutrinos themselves, and makes the problem non-linear. 
Interesting features of the oscillated neutrino energy spectra, caused by so-called spectral swaps, have been obtained by pioneering CNO studies  \cite{Duan2006CoherentEnvironment,Duan2006CollectiveSupernovae,Fogli2009SupernovaEffects}.
To predict the spectra,
Duan et al.~2006 formulated the basic quantities and oscillation modes \cite{Duan2006CollectiveSupernovae}.
Fogli et al.~2009 argued that spectral splits emerge
as dominant features in the inverted mass hierarchy \cite{Fogli2009SupernovaEffects}.
Note that the value of $\theta_{13}$ was not known very well at the time of these studies \cite{Ling2013}.

Although initial works on CNO assumed generic matter and neutrino profiles, recent works employ the results of numerical simulation of neutrino radiation hydrodynamics \cite{Chakraborty2011NoPhase,Chakraborty2011AnalysisPhase,Suwa2011ImpactsExplosions,Saviano2012StabilityDistributions,Sarikas2012SuppressionPhase,Wu2015EffectsModel,Dasgupta2012RoleRevival}. 
One important finding in these works is the importance of matter-induced decoherence, so-called matter suppression \cite{Chakraborty2011NoPhase,Chakraborty2011AnalysisPhase,Saviano2012StabilityDistributions,Sasaki2017PossibleProcesses,Zaizen:2018wfg}.
Essentially, if the matter density is higher than the neutrino density,
the wavelength of the neutrino's wave function becomes short and the
different paths from the proto neutron star cause significant decoherence of 
the wave functions. As a result, CNO is highly suppressed in such epochs. 
After this finding,
extensive studies of CNO were provided by Wu et al.~2015, where a progenitor of 18.0\Msun was used \cite{Wu2015EffectsModel}.
The rate of neutrino events expected at SK and DUNE were computed, and the possibility to distinguish the neutrino mass hierarchy has been suggested.

One important lesson from previous studies is that the final neutrino spectrum is significantly affected by not only the neutrino fluxes, but average energies and angular distributions.
Recently, Horowitz et al.~2017 proposed a new reaction rate of 
neutrino nucleon scattering
\cite{Horowitz2017Neutrino-nucleonExpansion}.
This causes the neutrino flux to significantly change, potentially impacting previous CNO results \cite{Wu2015EffectsModel}.

In this paper, we investigate the impact of 
CNO on the neutrino events from a nearby supernova and discuss its detectability.
We perform the first such study using the set of new neutrino reaction rates of Horowitz et al.~2017 and Horowitz 2002 \cite{Horowitz2017Neutrino-nucleonExpansion,Horowitz2002WeakSupernovae} as well as the other new reaction rates that are summarized in Ref. \cite{Kotake2018ImpactSimulations}.
These new reactions can change the neutrino emission and hence the resulting neutrino oscillation and detection. 
The structure of the paper is as follows.
In Section \ref{sec:methods}, we introduce our numerical schemes for the hydrodynamic simulations and the neutrino oscillation computations.
In Section \ref{sec:results}, we discuss our results, starting with the dynamics of the explosion, followed by features of neutrino oscillations, and finally, the detectability of CNO. In Section \ref{sec:summary}, we summarize our results.

\section{Methods}\label{sec:methods}
We performed two kinds of numerical simulations.
First one is the hydrodynamic simulation of core-collapse supernova from a progenitor model.
Second one is that for the three flavor neutrino oscillations using snapshots obtained by the hydrodynamic simulation.
In this section, we explain the numerical methods and settings for each. 

The hydrodynamic simulation was performed by {\bf 3DnSNe} code 
(see the references \cite{OConnor2018GlobalSymmetry,Kotake2018ImpactSimulations,Sasaki2017PossibleProcesses,Takiwaki2016Three-dimensionalFlows,Nakamura2016MultimessengerStrategies,Sotani2016GravitationalStars,Nakamura2015SystematicProgenitors} 
for recent application of this code).
The evolution of the variables are solved in coordinate of spherical polar geometry.
A piecewise linear method with the geometrical correction is used to reconstruct variables at the cell edge, where a modified van Leer limiter is employed to satisfy the condition of total variation diminishing (TVD) \cite{Mignone2014High-orderCoordinates}.
The numerical flux is calculated by HLLC solver \citep{Toro1994RestorationSolver}.
The models are computed on 1 dimensional spherical polar coordinate grid with a resolution of 512 radial zones.
The radial grid is logarithmically spaced and covers 
from the center up to the outer boundary of 5000 km.
Recently we updated our neutrino reactions \cite{Kotake2018ImpactSimulations}.
Among them, the effect of virial expansion on the neutrino nucleon scattering is important and this significantly changes the neutrino flux \cite{Horowitz2017Neutrino-nucleonExpansion}.
The equation of state used in the simulation is the Lattimer and Swesty with  incomprehensibility of  $K=220\ {\rm MeV}$ \cite{Lattimer1991AMatter}.
Although the code employs the relatively simple neutrino transport scheme of IDSA (Isotropic Diffusion Source Approximation) \cite{Liebendorfer2009TheTransport}, it nevertheless can provide consistent results on neutrino luminosities and average energies with more sophisticated schemes (see Ref.~\cite{OConnor2018GlobalSymmetry} for detailed comparison).

The progenitor employed in this study is a 8.8${\rm M}_\odot$ star \cite{Nomoto1984EvolutionCores,Nomoto1987EvolutionCore,Tominaga2013SUPERNOVASUPERNOVAE}.
The setup of the envelope is same to that of Kitaura et al.~2006 \cite{Kitaura2006ExplosionsSupernovae}.
Since the density of the envelope is low in this model,
the matter suppression is weak and 
the signature of CNO is expected. 
In the context of neutrino oscillation,
Saviano et al.~2012 also uses this progenitor \cite{Saviano2012StabilityDistributions}.

Flavor transitions of free-streaming neutrinos are calculated as post processes of the hydrodynamic simulation whose time snapshots give us the strength of the electron matter potential \cite{Wolfenstein1978NeutrinoMatter,Mikheev:1986gs} and that of neutrino self interaction \cite{Duan2006CollectiveSupernovae,Fogli:2007bk,Dasgupta:2009mg,Dasgupta:2010cd,Mirizzi:2010uz,Duan:2010bf,Wu2015EffectsModel,Sasaki2017PossibleProcesses,Birol:2018qhx,Zaizen:2018wfg}. Neutrino oscillation parameters in our simulation are given by the following values: $\sin^{2}(2\theta_{12})=0.84$, $\sin^{2}(2\theta_{23})=1$, $\sin^{2}(2\theta_{13})=0.19$, $\Delta m_{21}^{2}=7.9\times10^{-5}\ \mathrm{eV}^{2}$, $|\Delta m^{2}_{32}|=2.0\times10^{-3}\ \mathrm{eV}^{2}$ and $\delta_{\rm CP}=0$ where $\Delta m_{ij}^{2}=m_{i}^{2}-m_{j}^{2}$. We employ the bulb model \cite{Duan2006CollectiveSupernovae} and solve the time evolution of neutrino and anti-neutrino density matrices in three-flavor multiangle calculations based on Ref.~\cite{Sasaki2017PossibleProcesses}. 

We sample 50 points in neutrino energy $E\leq60$ MeV and choose $1000$ neutrino angular modes which are typical values to prevent numerical multiangle decoherence \cite{EstebanPretel:2007ec}.  
The radius of the neutrinosphere is fixed at 30 km which is close to sharp declines in the baryon density profiles.
In reality, the position of the neutrinosphere depends on the neutrino energy and flavor.
In the bulb model, multiple neutrinospheres however cannot be employed.
This is one of the major caveats of this study. 
The results of adopting a different neutrinosphere radius are shown in Appendix \ref{sec:neusphere}.
 
The spectra of supernova neutrinos often show pinched shapes
\cite{Tamborra2014NeutrinoSimulations,Tamborra2012High-resolutionFit}
compared to non-degenerate Fermi-Dirac distributions.
To capture this feature,
we set the initial neutrino spectra on the surface of the neutrinosphere,
$\phi_{i}(E)\ (i=\nu_{e},\bar{\nu}_{e},\nu_{X})$ as 
Gamma distribution, 
\begin{equation}
\phi_{i}(E) = \Phi_{i}^0
\frac{E^{\alpha_i}}{\Gamma_{\alpha_i+1}}
\left(\frac{\alpha_i+1}{\langle E_i \rangle}\right)^{\alpha_i+1}
\exp\left[-\frac{(\alpha_i+1)E}{\langle E_i\rangle }\right],\label{eq:gamma}
\end{equation}
whose pinching parameter, $\alpha_i=\frac{\langle E_i^2\rangle -2 \langle E_i\rangle^2}{\langle E_i\rangle^2-\langle E_i^2 \rangle}$, is estimated from the neutrino mean average energies $\langle E_i \rangle$ and rms average energies,$\sqrt{\langle E_i^2\rangle}$.
These parameters as well as the number luminosity, $\Phi_{i}^0$, are derived from the result of the neutrino radiation hydrodynamic simulation (see Section \ref{subsec:hydro} and the middle panel of FIG.~\ref{fig:t-le}). The multiangle calculations are carried out up to $1500$ km where CNO have finished. In our simulation, MSW resonances do not appear in neutrino energy $E\geq3$ MeV within $1500$ km because of high electron number density.
\section{Results}\label{sec:results}

We discuss the impact of neutrino oscillations on the detection rates at observation facilities HK, JUNO and DUNE.
We first introduce our hydrodynamic setups. In particular, the radial profile of the density and electron fraction as well as neutrino luminosity and energies are employed from the hydrodynamic simulation. Then, we introduce our calculation of neutrino oscillations, followed by detection. 

\begin{figure}[htbp]
\includegraphics[width=0.98\linewidth]{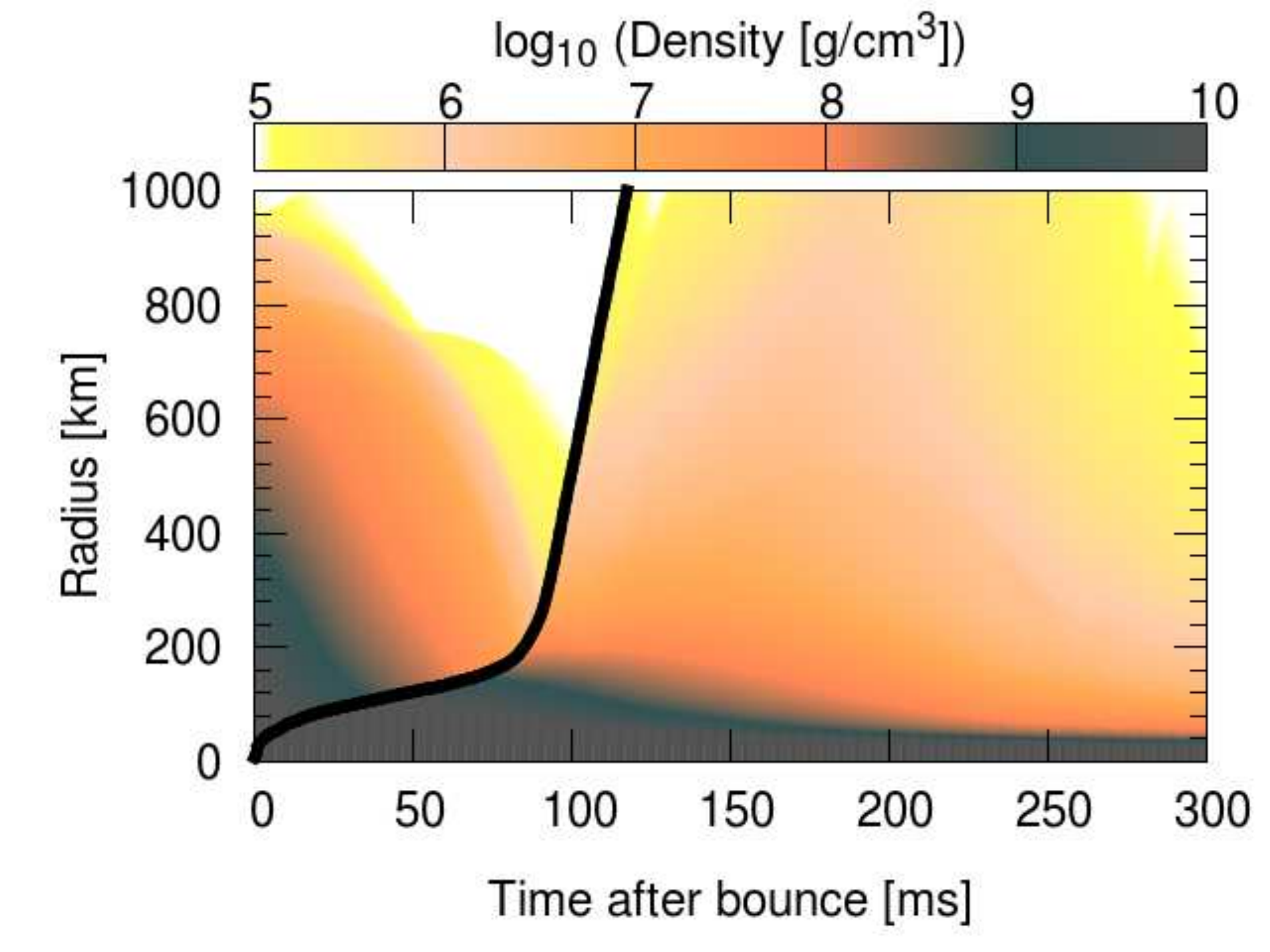}
\caption{
Black Curve: Time evolution of Shock Radius.
Color: Time evolution of logarithmic density profile ${\rm [g/cm^3]}$.
The horizontal axis is the time after bounce in ms and
the vertical axis is the radial coordinates in km. 
}
\label{fig:t-rs}
\end{figure}

\subsection{Hydrodynamic Model}\label{subsec:hydro}

The dynamics of the supernova explosion is characterized by the shock.
In our mode, the shock revives quite early by neutrino heating. 
The black curve of FIG.~\ref{fig:t-rs} shows the evolution of the averaged shock radius.
The shock revival time is 90 ms after bounce.
We adopt the widely used convention of shock revival time defined as the time when the shock reaches 400 km \cite{Summa2016PROGENITOR-DEPENDENTSUPERNOVAE}. 
This early shock revival time is due to the low mass accretion rates of this progenitor, which has very a diluted envelop (see FIG.~2 of Ref.~\cite{Yoshida2017ExplosiveElements}).
After shock revival, the shock continuously expands and reaches 1000 km at 120 ms after bounce. 
This result agrees with previous works (e.g.~FIG 3 of Ref.~\cite{Janka2008DynamicsSupernovae}). 
It should be noted that shock revival happens even in 1D geometry for this progenitor.

\begin{figure}[htbp]
\includegraphics[width=0.9\linewidth]{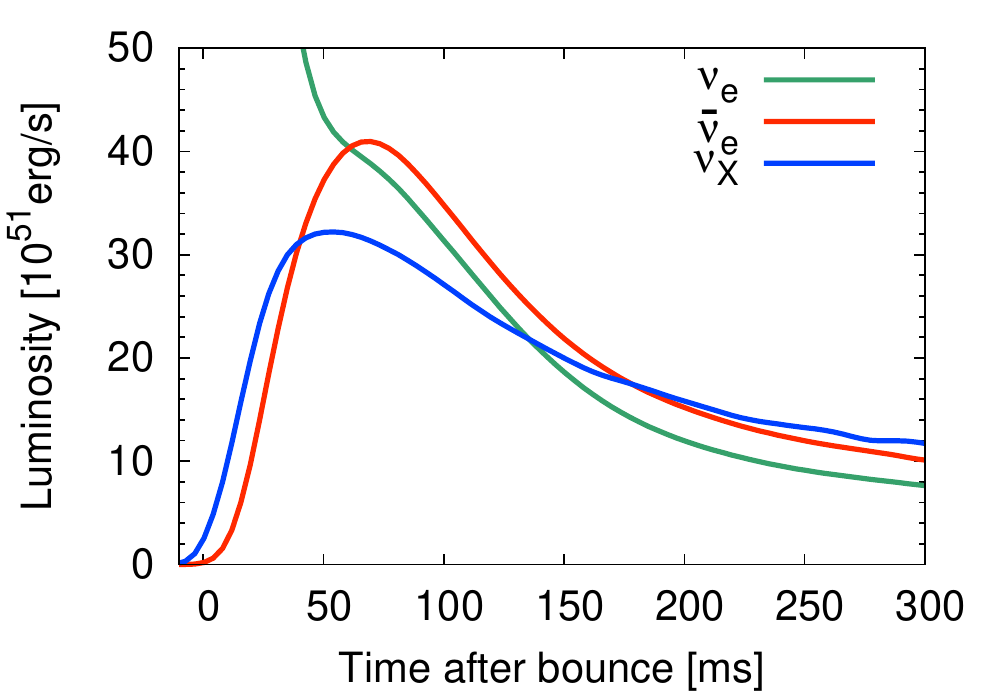}\\
\includegraphics[width=0.9\linewidth]{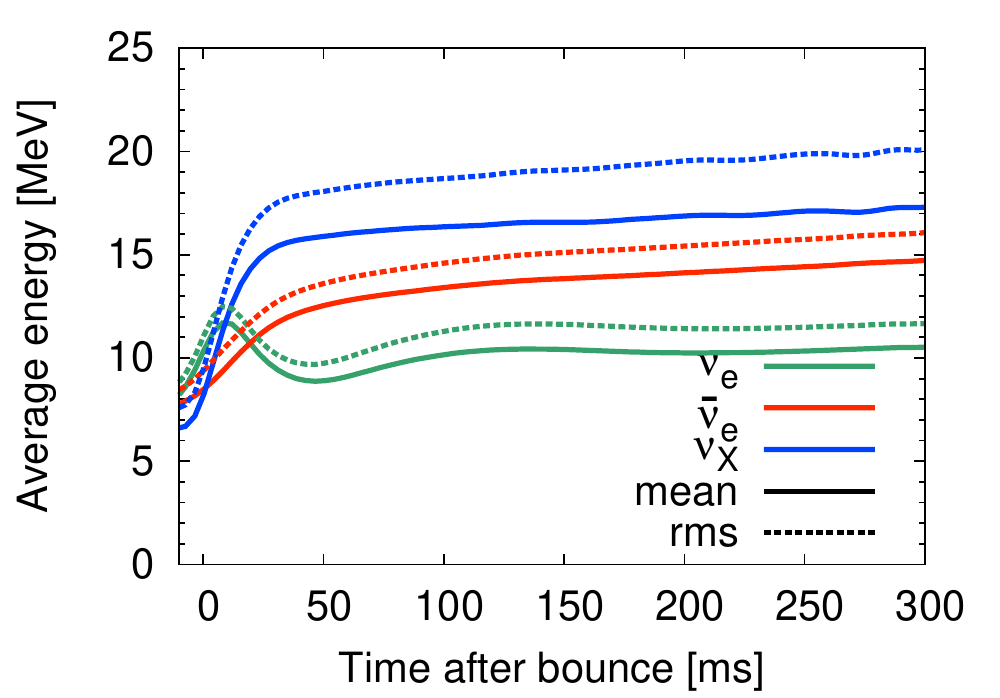}\\
\includegraphics[width=0.9\linewidth]{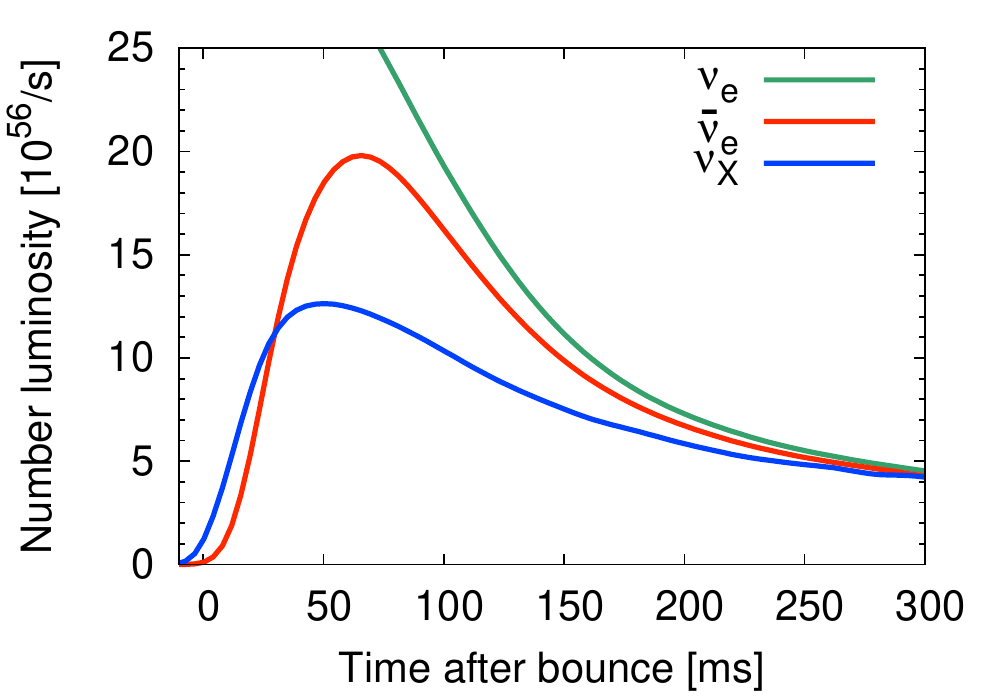}
\caption{
Time evolution of neutrino luminosity (top panel), neutrino average energy (central panel), and neutrino number luminosity (bottom panel). In all, the green, red, and blue curves correspond to $\nu_e$, $\bar{\nu}_e$ and $\nu_X$, respectively.
In the middle panel, the solid line denotes the mean energy and the dotted line the rms energy.
We evaluate the luminosities and energies at 500km and assume the bulb model, i.e.,
those luminosities and energies are set to the neutrino 
sphere in the oscillation simulations.
}
\label{fig:t-le}
\end{figure}

After shock revival, the density of the shocked region decreases.
The color map of FIG.~\ref{fig:t-rs} shows the logarithmic density as a function of time and radius.
As shown later, the region above 200 km is important for the neutrino oscillation in this model.
Before 50 ms postbounce, the density in the region is high and exceeds $10^{9}\ {\rm g/cm^3}$.
During this phase, the density gradually decreases with time due to the decrease of the mass accretion rate.
After the shock revival around 90 ms post bounce, the density briefly increases as a function of time since mass is ejected from the central region.
During this epoch, the density reaches $\sim 10^{7-8}\ {\rm g/cm^3}$ at $\sim 180$ ms post bounce.
However, after this the density decreases due to the decrease of mass ejection from the central region.
In the calculation of neutrino oscillation,
we select several time snapshots and 
use the profiles of density, $Y_e$ at the corresponding times.

The information of neutrino spectra is necessary for the input of the simulation of CNO.
The evolutions of the neutrino luminosities and energies are shown in FIG.~\ref{fig:t-le}.
The green, red and blue curves correspond to $\nu_e$, $\bar{\nu}_e$ and $\nu_X$, respectively.

In the top panel,
the luminosities after 150 ms post bounce are not so deviated from that of Ref.~\cite{Chakraborty2011AnalysisPhase}.
Before 150 ms, our luminosities are higher than that of Ref.~\cite{Chakraborty2011AnalysisPhase}
since an updated set of neutrino opacities is used (see FIG.~15 of Ref.~\cite{Kotake2018ImpactSimulations}).
In our model, the luminosity of anti-electron neutrino is larger than that of electron type neutrino.
This feature is not prominent in previous works (see FIG.~1 of Ref.~\cite{Chakraborty2011AnalysisPhase} and the hydrodynamic model of Ref.~\cite{Fischer2010ProtoneutronSimulations} for the detail of the setting).
Our feature may originate from the employment of weak magnetism \cite{Horowitz2002WeakSupernovae} that is not used in previous works.
The weak magnetism decreases the opacity for the anti-electron neutrino making them easier to escape. This would enlarge the anti-neutrino luminosity.

In the middle panel,
the hierarchy of the average energy is consistent with other simulations during the accretion phase:
$\nu_X > \bar{\nu}_e > \nu_e$.
The average is also higher compared to that of Ref.~\cite{Chakraborty2011AnalysisPhase} due to the new reaction set
 (see FIG.~15 of Ref \cite{Kotake2018ImpactSimulations} again).
The hierarchy of number luminosity has an interesting feature.

In the bottom panel, the number luminosity is shown and 
initially the hierarchy is 
$\nu_X < \bar{\nu}_e < \nu_e$, which is typical in the accretion phase of core-collapse supernovae. However, at 300 ms post bounce,
all number luminosities converge and there is no hierarchy.
This feature leads to interesting flavor mixing as discussed later.

\subsection{Neutrino Oscillation}
Using snapshots taken from the hydrodynamic simulation,
we calculated the neutrino oscillation.
We use rotated frame of $e-x-y$ instead of the flavor frame of $e-\mu-\tau$ \cite{Dasgupta2008CollectiveNeutrinos}.
In this section, we mainly focus on oscillations. The application to detection is discussed in the next section.

\subsubsection{Appearance of CNO}

\begin{figure}[htbp]
\includegraphics[width=0.95\linewidth]{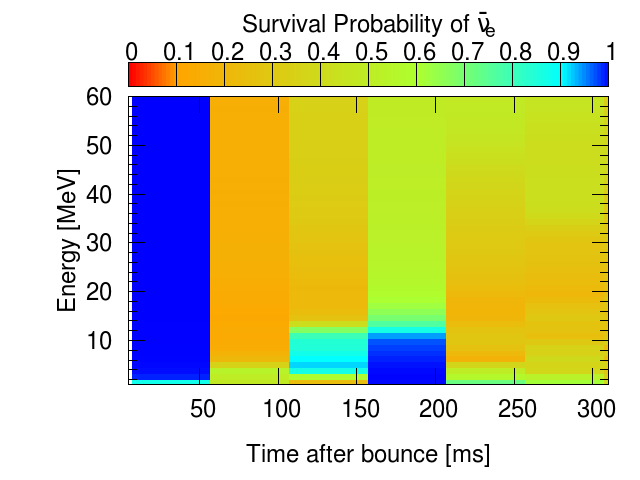}\\
\includegraphics[width=0.95\linewidth]{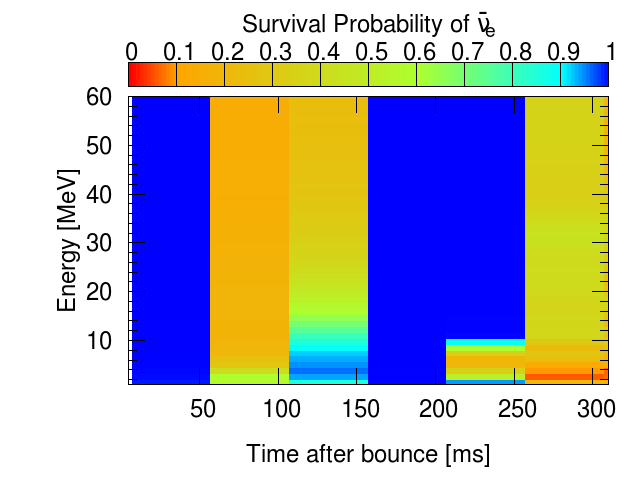}
\caption{
The survival probabilities of electron anti-neutrino 
at a radius of 1500 km as functions of neutrino energy 
and emission time. 
Top: the probability for the inverted mass hierarchy.
Bottom: the probability for the normal mass hierarchy.
In the both panels, we employ multi-angle method to 
calculate neutrino oscillations and the Gamma function 
(Eq.~\ref{eq:gamma}) is used for the neutrino spectrum.
}
\label{fig:T-E_SUV_G}
\end{figure}

The appearance of CNO is strongly affected by the density profile.
In our models, CNO appears before 100 ms post bounce.
In FIG.~\ref{fig:T-E_SUV_G}, the survival probability after CNO is shown as a function of the neutrino energy. The top panel is for the inverted mass hierarchy.
At 31 ms post bounce, the survival probability is almost $1.0$ and CNO is suppressed by the dense matter above the shock wave (the radius is above $\sim 200\ {\rm km}$).
At 81 ms post bounce, the density in the region becomes smaller and 
the survival probability becomes $0.2$ for most of the energy range except a narrow range of a few MeV.
The appearance of CNO in such an early epoch is due to our choice of the progenitor.
The 8.8 \Msun star has an extraordinary dilute envelop among the progenitors of core-collapse supernovae (see FIG.~1 of Ref. \cite{Janka2008DynamicsSupernovae}).
At 181 ms, the density of the region becomes slightly higher due to the shocked matter arising from the center. At this point, the survival probability becomes $1.0$ due to matter suppression, but only for neutrino energies below 10 MeV. The time of the occurrence of CNO also depends on the neutrino spectral shape.
If the spectral shape is more like a Fermi-Dirac, the onset time is significantly delayed and becomes 231 ms (see FIG.~\ref{fig:T-E_SUV_FD} in Appendix \ref{sec:FD}).

The survival probability in the case for the normal mass hierarchy is roughly similar to that of the inverted mass hierarchy, but generally more suppressed. 
The bottom panel of the FIG.~\ref{fig:T-E_SUV_G} shows the evolution as functions of neutrino energy and time. Like in the inverted mass hierarchy, the appearance of CNO occurs before $ 100\ {\rm ms}$.
However, at 181 ms postbounce, the CNO is suppressed due to the increase in the matter density over all neutrino energies (see the density evolution at $\sim 200\ {\rm  km}$ in FIG.~\ref{fig:t-rs}). Though the suppression also appears in the inverted mass hierarchy, it is stronger in the normal mass hierarchy case.
A strong suppression is also seen in the Fermi-Dirac neutrino spectrum (see FIG.~ \ref{fig:T-E_SUV_FD} in Appendix \ref{sec:FD}), where it can be seen that the appearance of CNO is delayed to 281 ms which is later than in the inverted mass hierarchy, 231 ms.

\subsubsection{Neutrino oscillation in the inverted mass hierarchy}
\begin{figure}[htbp]
\includegraphics[width=0.95\linewidth]{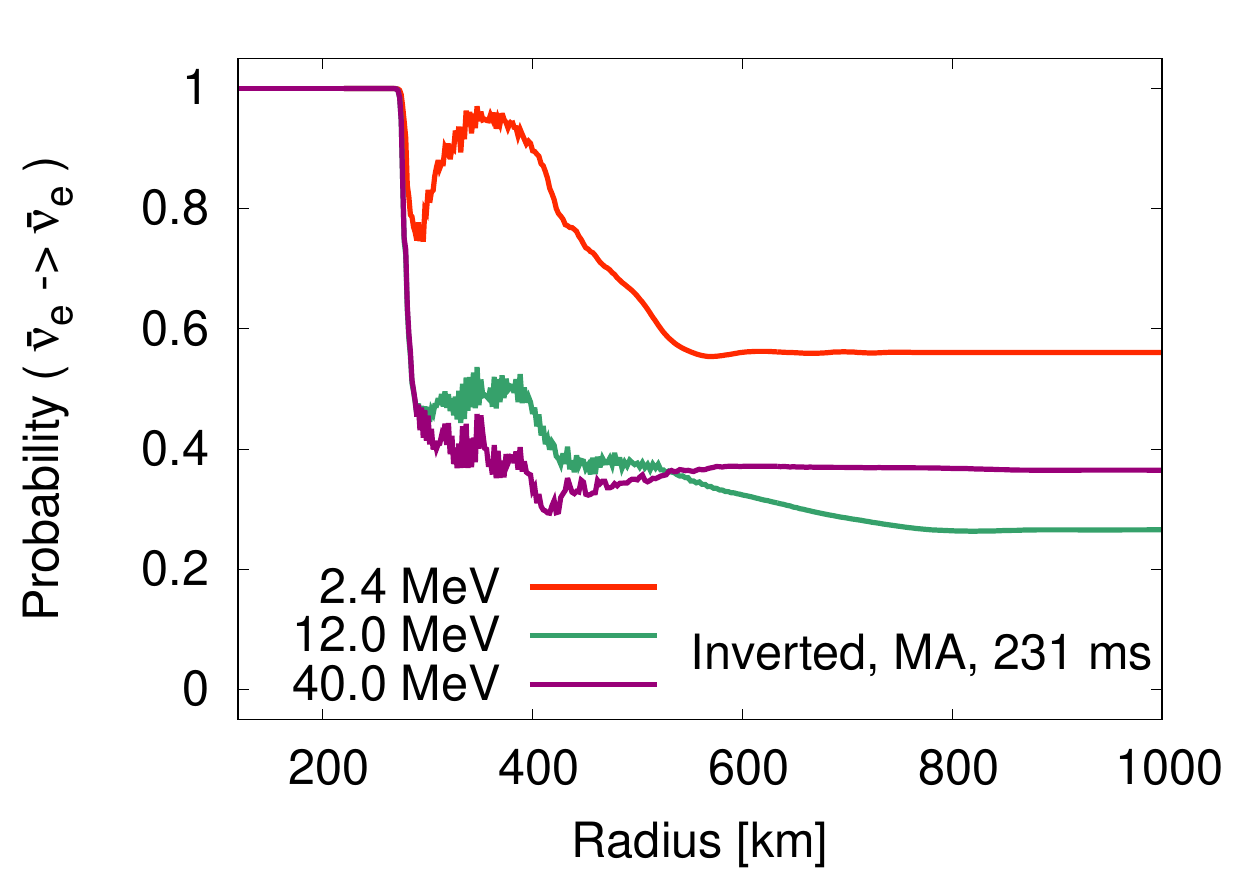}
\includegraphics[width=0.95\linewidth]{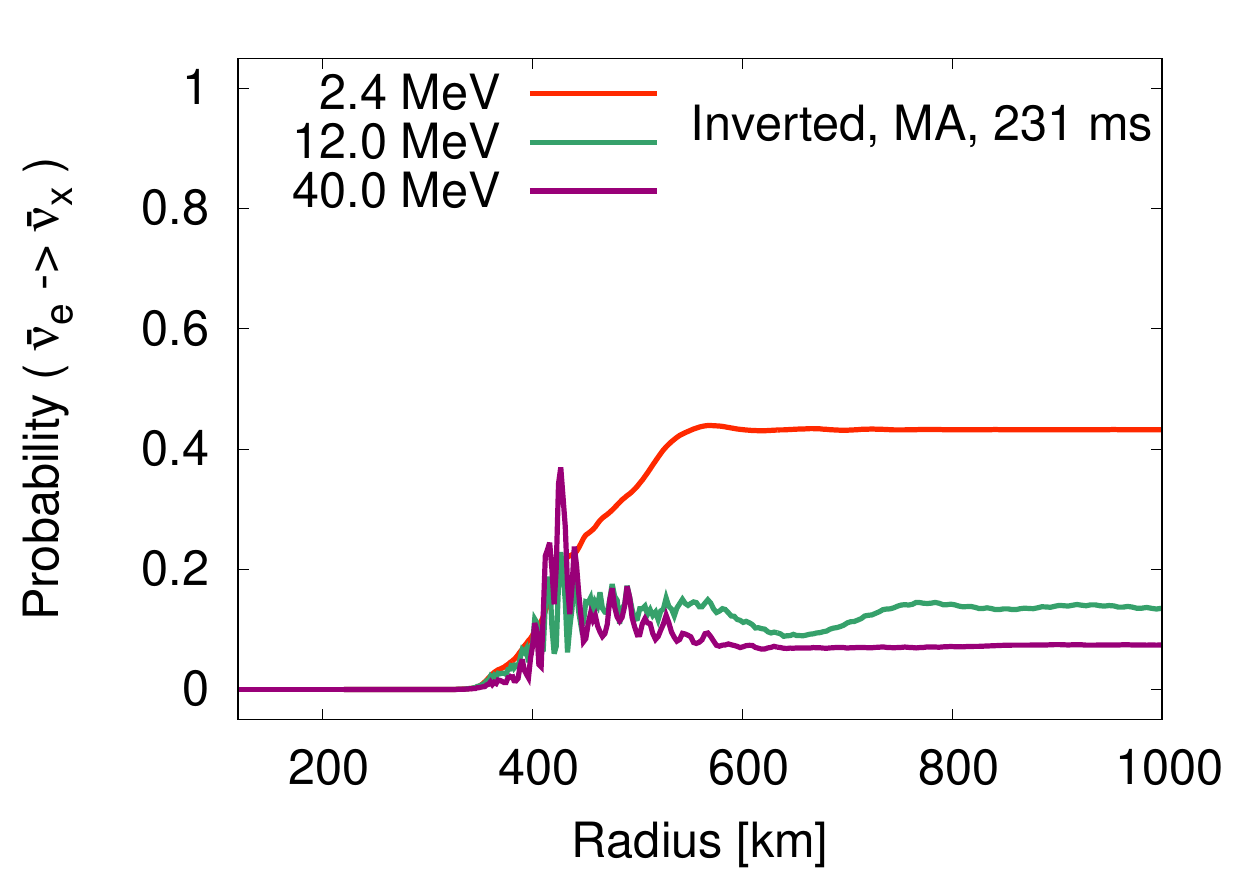}
\includegraphics[width=0.95\linewidth]{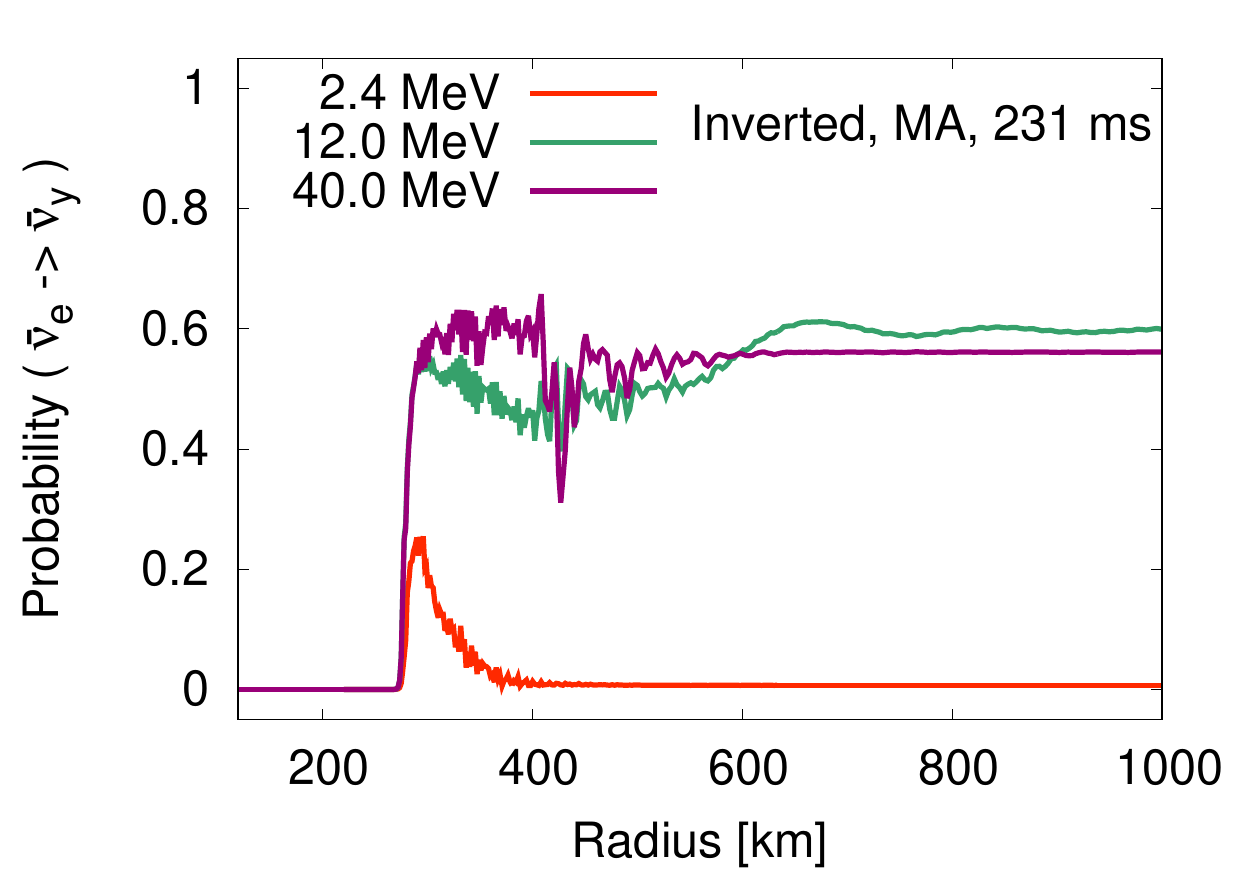}
\caption{
The radial profiles of conversion probabilities of $\bar{\nu}_{e}$ at $231\ {\rm ms}$ post bounce. 
The inverted mass hierarchy is assumed and multiangle scheme (labeled MA) is used.
The different colors show the profile at different energies of the neutrino:
red, green and violet correspond to 2.4, 12.0 and 40.0 MeV, respectively.
}
\label{fig:r-suv}
\end{figure}

The radial profiles of conversion probabilities $P_{e\alpha}=P(\bar{\nu}_{e}\to\bar{\nu}_{\alpha})\ (\alpha=e,x,y)$ \cite{Dasgupta:2010cd} are helpful to understand the behavior of non-linear flavor transitions. In FIG.~\ref{fig:r-suv}, we present the conversion probabilities of anti-neutrinos at $231$ ms in the inverted mass hierarchy. Such probabilities are derived from angle averaged diagonal components of density matrices (see Eq.~(11) in Ref.~\cite{Sasaki2017PossibleProcesses}). The top panel corresponds to the evolution of survival probabilities of $\bar{\nu}_{e}$. The value of $P_{ee}$ remains unity as long as flavor transitions are negligible. The middle and bottom panels show how emitted $\bar{\nu}_{e}$ on the surface of the neutrinosphere is transformed to $\bar{\nu}_{y}$ and  $\bar{\nu}_{x}$, respectively.
We are solving the time evolution of neutrino and anti-neutrino density matrices in three-flavor multiangle calculations.
Non-diagonal components in neutrino self interactions Hamiltonian grow up prominently when once self interactions term couple with vacuum Hamiltonian. Such non-diagonal potential gives rise to non-linear collective motion in flavor space.

As expected, the electron anti-neutrino essentially experience CNO 
in the $e-y$ sector in the inverted mass hierarchy. In FIG.~\ref{fig:r-suv}, 
we show the flavor evolution in the inverted mass hierarchy for three 
$\bar{\nu}_e$ energies: 2.4 MeV, 12.0 MeV, and 40.0 MeV. It can be seen 
that CNO starts around $250$ km in the 
$e-y$ sector: the survival probability decreases (top panel) and the 
conversion probability for $e-y$ becomes larger (bottom panel), except for the low energy component of 2.4 MeV.

Subsequent CNO occurs in $\bar{\nu}_{x}$ at $450$ km after the early $e-y$ mixing. Two types of non-linear transitions reflect the coupling of self interaction with two vacuum frequencies, $\omega_{\mathrm{solar}}=\Delta m^{2}_{21}/2E$ and $\omega_{\mathrm{atm}}=|\Delta m^{2}_{32}|/2E$.  
Such three-flavor peculiar mixing is also found in previous numerical studies \cite{Dasgupta:2010cd,Mirizzi:2010uz} and arise from a small flavor asymmetry in the neutrino number luminosity, e.g., at $231$ ms: $\Phi^{0}_{\nu_{e}}:\Phi^{0}_{\bar{\nu}_{e}}:\Phi^{0}_{\nu_{x}}=1.17:1.09:1.00$. As shown in the bottom panel of FIG.~\ref{fig:t-le}, the flavor asymmetry becomes smaller as time proceeds, which enhances the three flavor mixing in the post-accretion phase. At $600$ km, CNO has finished and conversion probabilities settle down to constant values. Then final flavor mixing is energy dependent. For example, low energy electron anti-neutrinos transform to other flavor $\bar{\nu}_{x}$ actively and $\bar{\nu}_{y}$ returns to the original flux as shown by the $2.4$ MeV curves in FIG.~\ref{fig:r-suv}. On the other hand, $e-x$ mixings become small in more energetic electron anti-neutrinos, for example $12.0$ and $40.0$ MeV, even though the vacuum frequency $\omega_{\mathrm{atm}}$ induces partially $\bar{\nu_{y}}-\bar{\nu}_{x}$ conversions.

\begin{figure}[t]
\includegraphics[width=0.95\linewidth]{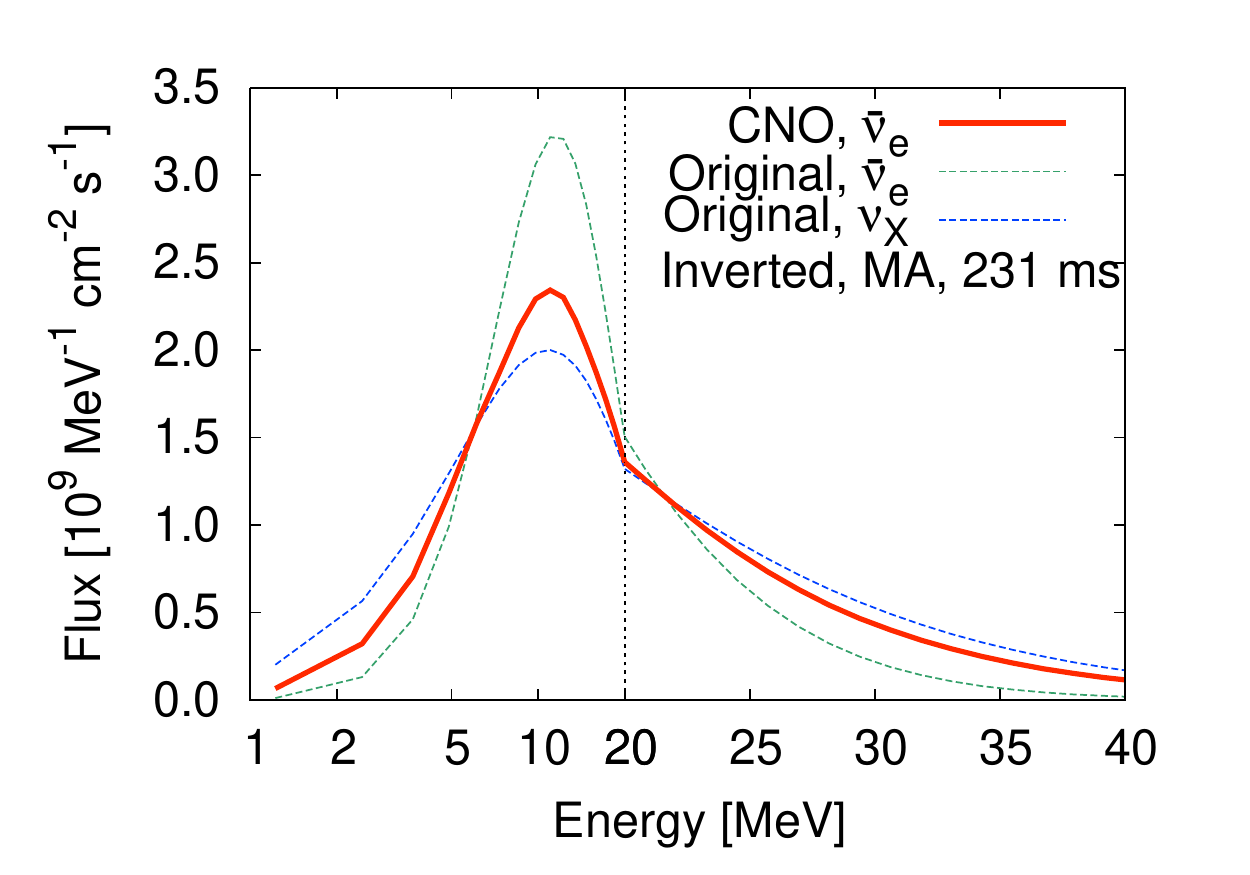}\\
\includegraphics[width=0.95\linewidth]{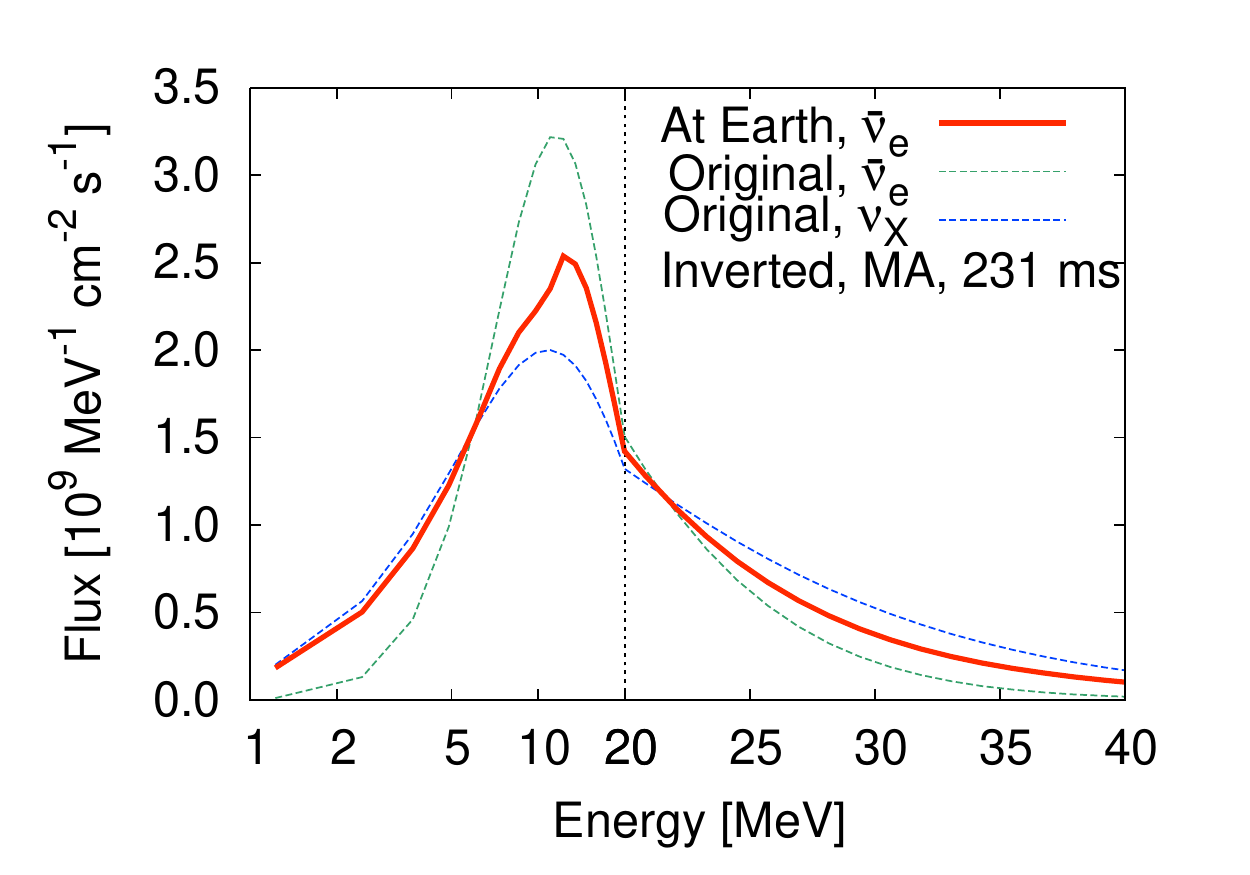}
\caption{
Flux spectra of electron anti-neutrinos in the inverted mass hierarchy (red solid). For comparison, the original electron neutrino (green dotted) and heavy lepton neutrino (blue dotted) are shown. The time snapshot of 231 ms is used.
Top panel: CNO with the MA scheme (1500 km from the center; note that the vertical axis has been rescaled to a source distance of 10 kpc). Bottom panel: flux spectrum at Earth, including all oscillation effects (CNO, MSW, vacuum). Note that in both panels, the horizontal axis is logarithmic for $E<20\ {\rm MeV}$ and liner for $E>20\ {\rm MeV}$ in the both panels (indicated by the vertical dashed line).
}
\label{fig:E-Fdetail}
\end{figure}

The top panel of FIG.~\ref{fig:E-Fdetail} shows the spectrum of anti-neutrinos at 231 ms post bounce after the CNO (i.e., at 1500km, in red), compared to the original $\bar{\nu}_e$ (green) and $\nu_X$ (blue). 
This $\bar{\nu}_{e}$ spectrum traces the property of survival probabilities in the top panel of FIG.~\ref{fig:r-suv}.
In the high energy range, $E>12\ {\rm MeV}$, a spectral oscillation occurs actively in the $e-y$ sector. The $e-x$ conversion is suppressed. Complete spectral swaps, which are generally observed under the single angle approximation \cite{Duan2006CollectiveSupernovae}, fails in multiangle simulations because the  coherence of non-linear flavor transitions are smeared out by angular dispersion in the matter potential \cite{EstebanPretel:2008ni}. 
In the region of $E<12\ {\rm MeV}$, the spectrum is rather closer to the original spectrum and
interestingly the $e-x$ conversion is dominant.
In general, three flavor mixing would induce complex spectral swaps in the outer layers different from a simple two flavor picture. However, the low energy events do not strongly contribute to the total event rates.

\begin{figure}[htbp]
\includegraphics[width=0.95\linewidth]{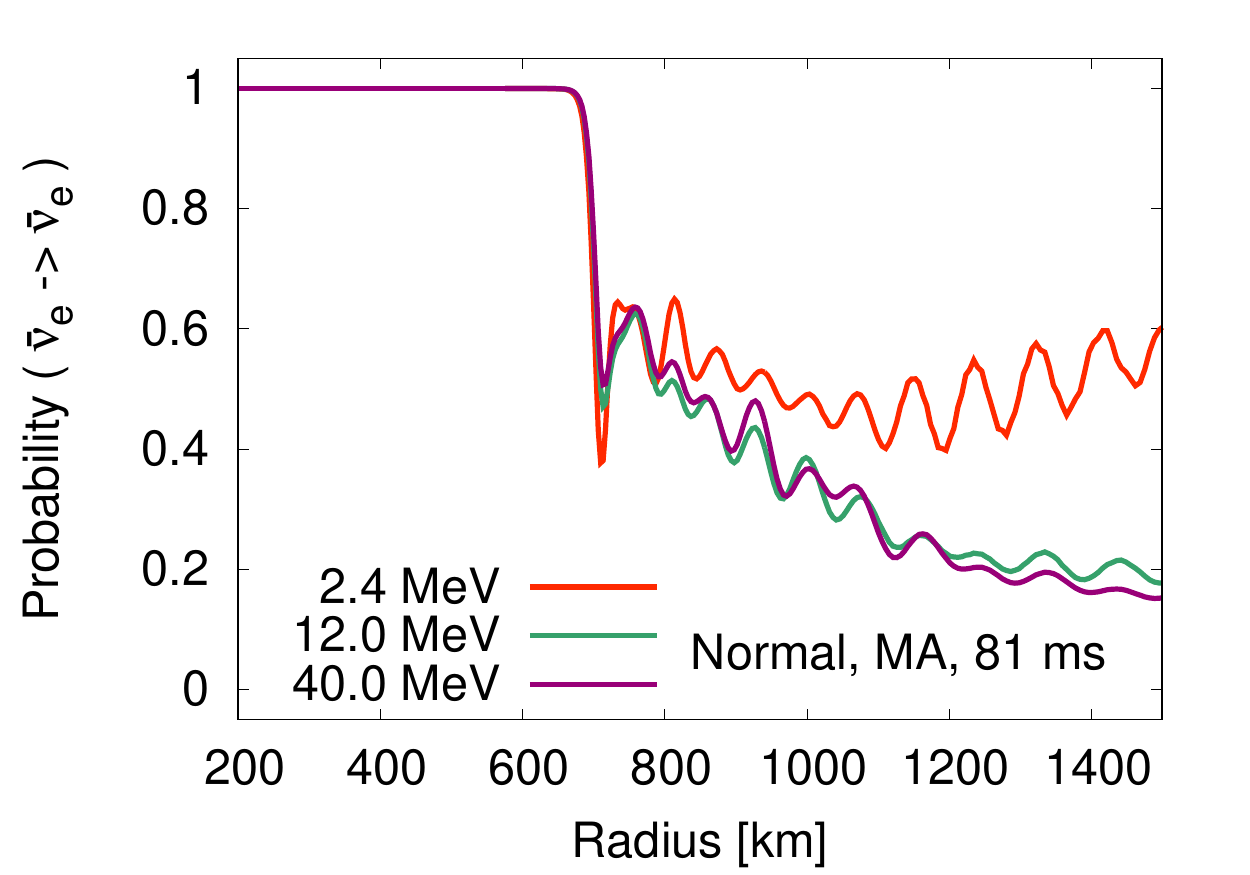}
\includegraphics[width=0.95\linewidth]{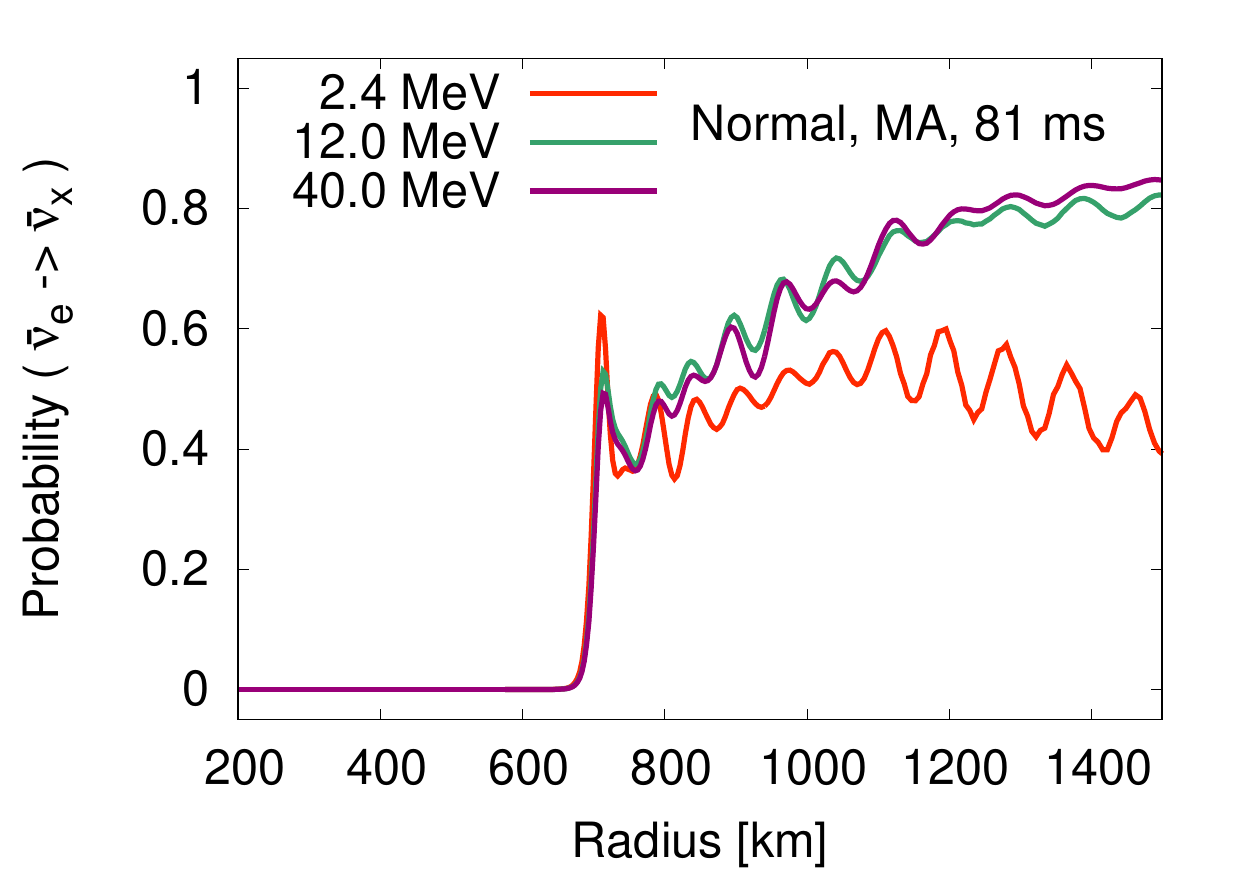}
\includegraphics[width=0.95\linewidth]{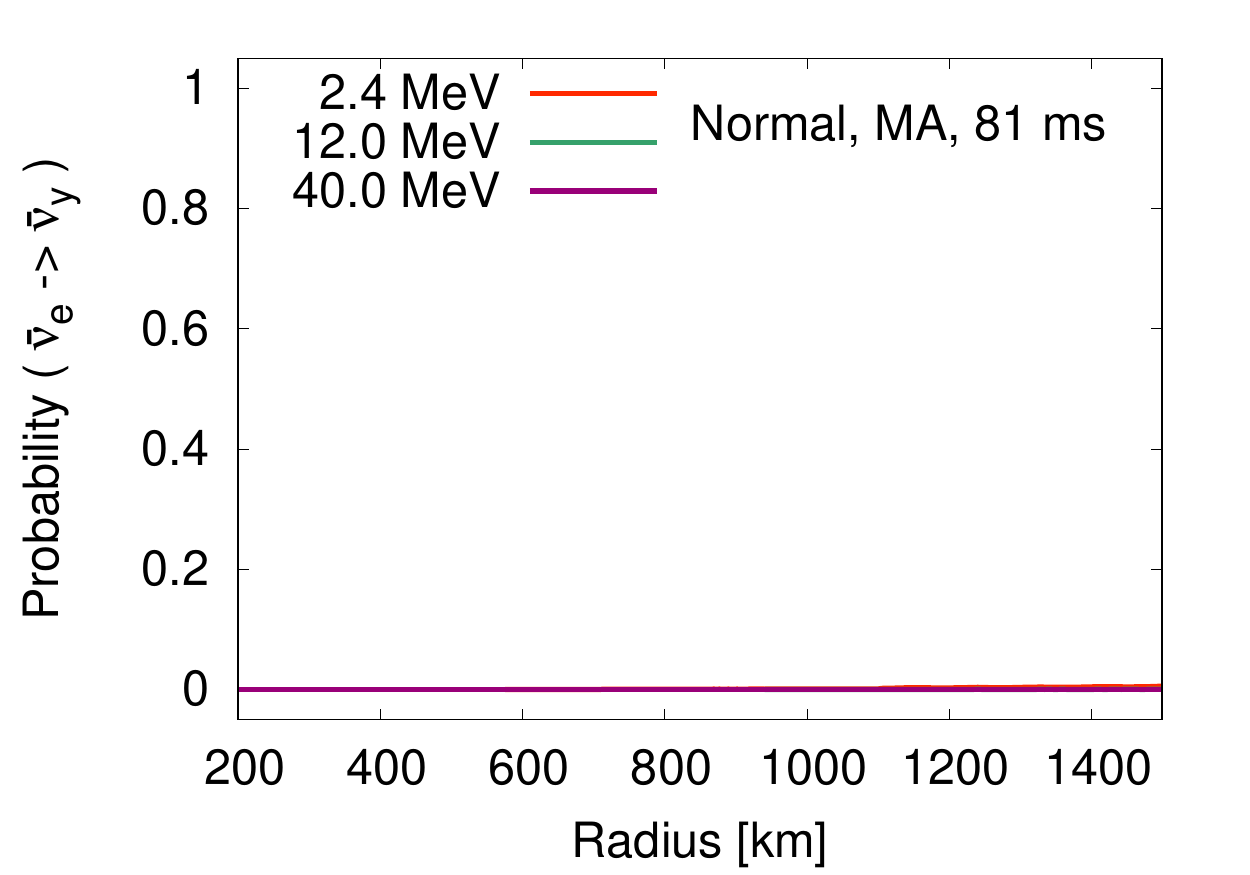}
\caption{
The radial profiles of conversion probabilities of $\bar{\nu}_{e}$ at $81\ {\rm ms}$ post bounce. 
The normal mass hierarchy is assumed and multiangle scheme (labeled MA) is used.
The different colors show the profile at different energies of the neutrino:
red, green and violet correspond to 2.4, 12.0 and 40.0 MeV, respectively.
}
\label{fig:r-suv_normal}
\end{figure}

\begin{figure}[htbp]
\includegraphics[width=0.95\linewidth]{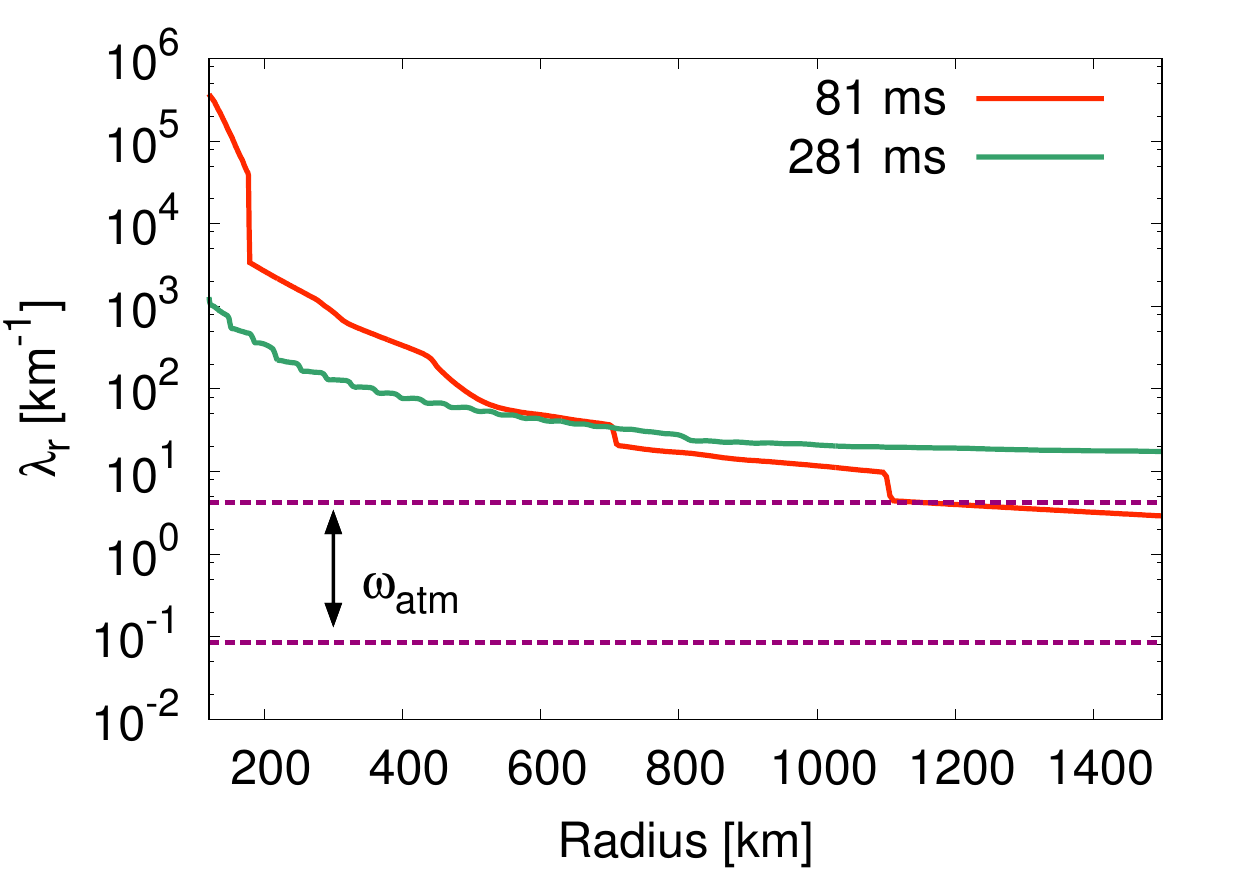}
\caption{
The time snapshots of matter potential $\lambda_{r}=\sqrt{2}G_{F}n_{e}$ for $81$ and $281$ ms post bounce. The horizontal band shows the values of atmospheric vacuum frequencies $\omega_{\mathrm{atm}}=|\Delta m^{2}_{32}|/2E$ for $E\ [\mathrm{MeV}]\in[1.2,60]$. The upper line corresponds that of the lowest energy. The MSW resonance occurs when the matter potential enters the band of $\omega_{\mathrm{atm}}$.
}
\label{fig:density_profile_332_2}
\end{figure}

\begin{figure}[htbp]
\includegraphics[width=0.95\linewidth]{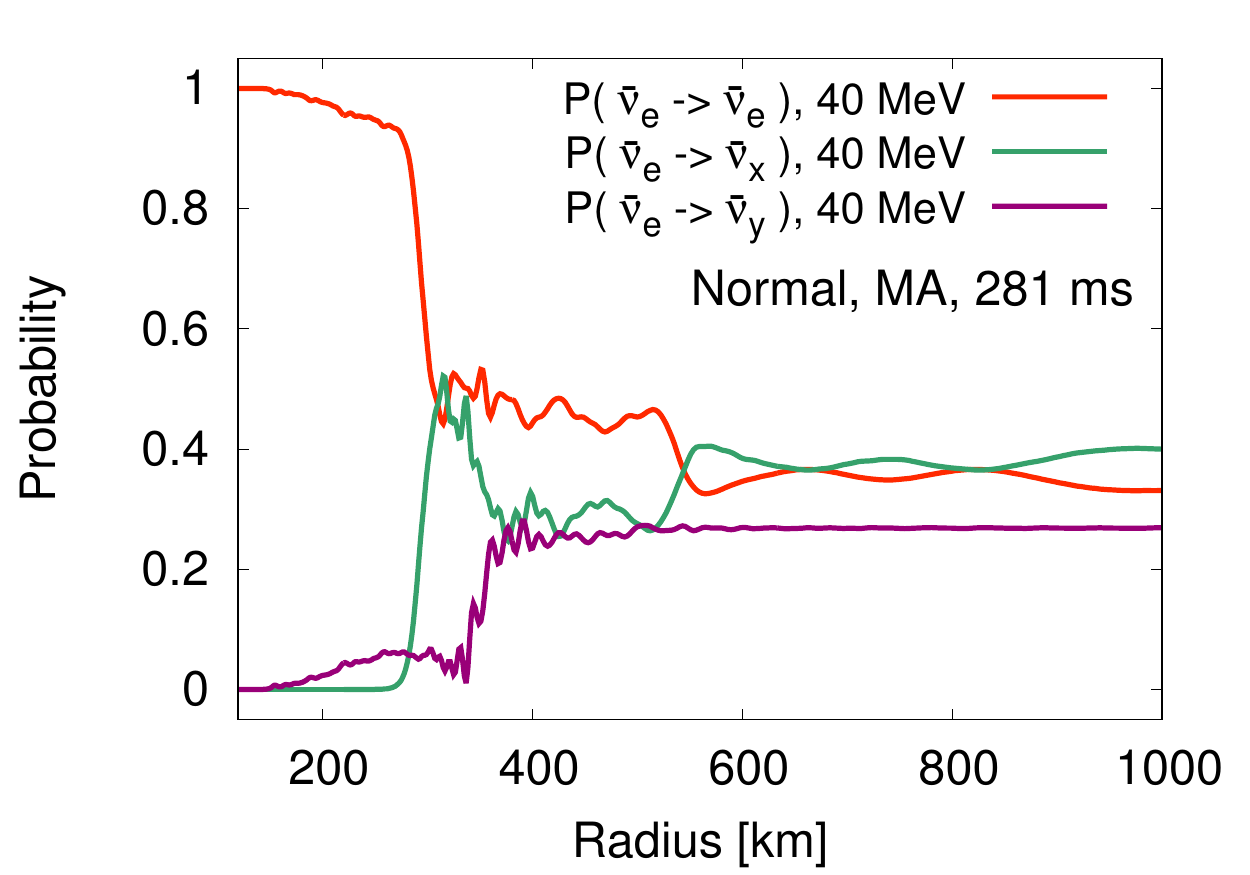}
\caption{
The radial profiles of conversion probabilities of electron antineutrinos at $281\ {\rm ms}$ post bounce in the normal mass hierarchy. The different colors show the profiles of different conversion probabilities of $\bar{\nu}_{e}$: $P_{e\alpha}=P(\bar{\nu}_{e}\to\bar{\nu}_{\alpha})\ (\alpha=e,x,y)$ whose energy is $40$ MeV.
The red, green and violet lines correspond to $P_{ee}$, $P_{ex}$ and $P_{ey}$, respectively.
}
\label{fig:r-suv_normal_332}
\end{figure}

\subsubsection{Neutrino oscillation in the normal mass hierarchy}
In the case of normal mass hierarchy, CNO appears in two typical time domains, $\sim 100\ {\rm ms}$ and $\sim 300\ {\rm ms}$ post bounce, while it is suppressed around $\sim 200\ {\rm ms}$. In the case of inverted mass hierarchy, CNO continues after 50 ms after bounce. 
Such hierarchy differences may come from the multiangle suppression \cite{Mirizzi:2010uz} of the  $e-x$ and $e-y$ sectors.

At $\sim 100\ {\rm ms}$, the $e-x$ conversion occurs dominantly in the normal mass hierarchy. The role of $\bar{\nu}_{x}$ in the normal mass hierarchy is that of $\bar{\nu}_{y}$ in the inverted mass hierarchy.
In the inverted mass hierarchy, the $e-y$ conversion occurs dominantly thorough almost all energy region and in only low energy region, $e-x$ conversion can happen. This feature is clearly seen in FIG. \ref{fig:r-suv_normal}. The survival probabilities of 12.0 MeV and 40.0 MeV tend to about $0.2$, while the survival probability of low energy neutrino (2.4 MeV) is about $0.5$ at 1500 km.
A significant fraction of electron anti-neutrino is converted to $x$ anti-neutrino. CNO takes place before the MSW resonances because the matter potential, $\lambda_{r}=\sqrt{2}G_{\rm F}n_{e}$, is higher than the atmospheric vacuum frequency, $\omega_{\mathrm{atm}}$, around the onset of CNO $\sim700$ km. The $\lambda_{r}$ and $\omega_{\mathrm{atm}}$ are shown in  FIG.\ref{fig:density_profile_332_2}.
The upper band of the $\omega_{\mathrm{atm}}$ comes from that of the lowest energy, 1.2 MeV.
The low energy neutrino ($E<3$ MeV) enters the region of MSW effects before the outer boundary of 1500km. Since the low energy neutrino does not strongly affect the observational signal, we do not focus on this effect.

At $\sim 200\ {\rm ms}$, the matter follows the expanding shock and the density above 200km becomes higher.
CNO ceases at that time by the strong matter suppression.

At $\sim 300\ {\rm ms}$, CNO again occurs because the electron density becomes lower outside the neutrinosphere. The three flavor mixing is significant due to the small asymmetry between neutrino number luminosities\cite{Mirizzi:2010uz}.
We find that several types of CNO contribute in different radius: $e-y$ mixing ($r<270\ {\rm km}$),
$e-x$ mixing ($r>270\ {\rm km}$) and
$x-y$ mixing ($r>340\ {\rm km}$) in high energy neutrinos and antineutrinos.
For example, such three flavor mixing of $40$ MeV antineutrinos at $281$ ms is shown in FIG.\ref{fig:r-suv_normal_332}. The small amount of $e-y$ conversion starts around $150$ km, which comes from the coupling of neutrino self interactions with $\omega_{\mathrm{atm}}$. After that, the significant $e-x$ conversion occurs around $270$ km in this later explosion phase. 
Then, a subsequent $x-y$ conversion around $340$ km increases the three flavor mixing in high energy antineutrinos. Such flavor conversion is also confirmed in neutrinos. It seems that, in the normal mass hierarchy, the $e-x$ conversion becomes main process of CNO in high energy neutrinos and antineutrinos across the examined time window irrespective of the three flavor mixing in the later explosion phase. FIG.\ref{fig:density_profile_332_2} indicates that the origin of the $e-x$ conversion in the normal mass hierarchy does not come from `` MSW prepared spectral split" as confirmed in the neutronization neutrino burst \cite{Dasgupta:2008cd,Duan:2007sh,Cherry:2010yc} because the value of $\lambda_{r}$ does not reach that of $\omega_{\mathrm{atm}}$ at the onset of the $e-x$ conversion $\sim270$ km (see the green curve of FIG.\ref{fig:density_profile_332_2}).

The significant $e-x$ conversion in the normal mass hierarchy has not been shown by previous three flavor simulations \cite{Dasgupta2008CollectiveNeutrinos,Dasgupta:2010cd,Mirizzi:2010uz}. We discuss a possible mechanism of such a new type of flavor mixing which would be equivalent to an instability for normal mass hierarchy suggested in a recent work \cite{Doring:2019axc}. We begin with a simpler problem. 
Namely, in two flavor collective neutrino oscillations, the relation between the direction of vacuum polarization vector $\omega \mathbf{B}$ and nonlinear polarization vector $\mathbf{D}$ is crucial for the development of nonlinear effects \cite{Hannestad:2006nj}. Here, the direction of polarization vector $\mathbf{P}=(P_{x},P_{y},P_{z})$ is given by the sign of the $z$-component (note in this discussion the meanings of $x,y$ are different from that of the rotated basis). The initial value of nonlinear potential is independent of neutrino mass hierarchy, so we only focus on the $z$-component of the vacuum polarization vector in each neutrino sector.
Now in reality, we have to consider three flavor cases. 
In the rotated basis $(\nu_{e},\nu_{x},\nu_{y})$, the vacuum Hamiltonian of three flavor neutrinos \cite{Sasaki2017PossibleProcesses} is described by
\begin{equation}
\begin{split}
\label{eq:3 flavor vacuum detail}
\Omega(E)
&=\frac{\Delta m_{21}^{2}}{6E}\left(
\begin{array}{c c c}
1-3c_{12}^{2}c_{13}^{2}&3c_{12}s_{12}c_{13}&3c_{12}^{2}s_{13}c_{13}\\
3c_{12}s_{12}c_{13}&1-3s_{12}^{2}&-3s_{12}c_{12}s_{13}\\
3c_{12}^{2}s_{13}c_{13}&-3s_{12}c_{12}s_{13}&1-3c_{12}^{2}s_{13}^{2}\\
\end{array}\right)\\
&+\frac{\Delta m_{32}^{2}}{6E}\left(
\begin{array}{c c c}
-1+3s_{13}^{2}&0&3s_{13}c_{13}\\
0&-1&0\\ 
3s_{13}c_{13}&0&-1+3c_{13}^{2}\\
\end{array}\right),
\end{split}
\end{equation} 
where $c_{ij}$ and $s_{ij}$ stand for $\cos\theta_{ij}$ and $\sin\theta_{ij}$, respectively. This three flavor vacuum Hamiltonian is equivalent to Eq.~($6$) in Ref.~\cite{Dasgupta2008CollectiveNeutrinos} even though we employ different normalization: $\mathrm{Tr}\ \Omega(E)=0$.
The nonlinear potential of neutrino self interactions first couples to the second term on the right hand side of Eq.~(\ref{eq:3 flavor vacuum detail}) because $|\Delta m^{2}_{32}|/\Delta m^{2}_{21}>O(10)$. From the definition of polarization (Bloch) vectors in a $2\times2$ Hermite matrix \cite{Fogli:2007bk}, the $z$-component of the vacuum Hamiltonian in the $e-y$ sector is obtained by the difference between two diagonal components in the $e-y$ sector,
\begin{equation}
\label{eq:relative sign of diagonal components e-y}
    \Omega(E)_{ee}-\Omega(E)_{yy}\sim -\frac{\Delta m_{32}^{2}}{2E}\cos2\theta_{13},
\end{equation}
where the sign of $\Delta m_{32}^{2}$ depends on the neutrino mass hierarchy. If the sign of $D_{z}$ is positive, $\Phi^{0}_{\nu_{e}}>\Phi^{0}_{\bar{\nu}_{e}}$, the positive sign in Eq.~(\ref{eq:relative sign of diagonal components e-y}) is preferable for significant flavor transitions in the $e-y$ sector as shown in two flavor CNO in the inverted mass hierarchy ($\Delta m_{32}^{2}<0$) \cite{Duan2006CollectiveSupernovae,Fogli:2007bk}. 
We remark that some $e-y$ conversion also appears even in the normal mass hierarchy (e.g., the slight increase of $P_{ey}$ within $270$ km in FIG.\ref{fig:r-suv_normal_332}) if the asymmetry among neutrino number luminosities $\Phi^{0}_{i}(i=\nu_{e},\bar{\nu}_{e},\nu_{x})$ is small enough to induce multiple spectral swaps in the inverted mass hierarchy \cite{Dasgupta:2010cd}. In our explosion model, the asymmetry of neutrino number luminosity gradually decreases as shown in the bottom panel of Fig.~\ref{fig:t-le}. The $e-x$ conversions in the normal mass hierarchy can be discussed in the same way as $e-y$ conversions above. Large instabilities would appear in $e-x$ conversions if the $z$-component of the vacuum Hamiltonian in the $e-x$ sector,

\begin{equation}
\begin{split}
\label{eq:relative sign of diagonal components e-x}
    \Omega(E)_{ee}-\Omega(E)_{xx}&=-\frac{\Delta m_{21}^{2}}{2E}\left(
    \cos2\theta_{12}-\cos^{2}\theta_{12}\sin^{2}\theta_{13}
    \right)\\
    &+\frac{\Delta m_{32}^{2}}{2E}\sin^{2}\theta_{13}.
    \end{split}
\end{equation}
takes positive values. In the inverted mass hierarchy $(\Delta m^{2}_{32}<0)$, the sign of Eq.~(\ref{eq:relative sign of diagonal components e-x}) is always negative because of $\Delta m^{2}_{21}>0$ and $|\Delta m^{2}_{32}|/\Delta m^{2}_{21}>O(10)$. However, Eq.~(\ref{eq:relative sign of diagonal components e-x}) becomes positive in the normal mass hierarchy $(\Delta m^{2}_{32}>0)$ if we impose a finite mixing angle $\theta_{13}$ larger than below a critical value
\begin{equation}
\label{eq:condition e-x mixing}
    \sin^{2}\theta_{13}>\frac{\cos2\theta_{12}}{\Delta m_{32}^{2}/\Delta m_{21}^{2}+\cos^{2}\theta_{12}}.
\end{equation}
This condition may be also applicable in a dense electron background before MSW resonances, because diagonal terms in the matter potential can be canceled out in a co-rotating frame \cite{Duan2006CollectiveSupernovae} which moves together with the non-linear potential of neutrino self interactions. Therefore, the above criterion may be applicable to a sparse electron background. Our neutrino mixing parameters satisfy the above condition. This is also true for more updated values of neutrino mixing parameters \cite{Tanabashi:2018oca}. On the other hand, Eq.~(\ref{eq:condition e-x mixing}) is violated in case of small mixing angle $\theta_{13}$ used in previous studies \cite{Dasgupta2008CollectiveNeutrinos,Dasgupta:2010cd,Mirizzi:2010uz}. This seems to be a plausible reason why the $e-x$ conversions in the normal mass hierarchy are discovered in our simulation but not confirmed in Refs.~\cite{Dasgupta2008CollectiveNeutrinos,Dasgupta:2010cd,Mirizzi:2010uz}. As shown in the bottom panel of FIG.\ref{fig:T-E_SUV_G}, flavor conversions in the normal mass hierarchy are easily suppressed in a dense electron background. Therefore, the $e-x$ conversions in the normal mass hierarchy would be fragile in more massive progenitor models. Nevertheless, further studies are necessary for more robust conclusions.

\subsection{Detectability of the feature of CNO}
Next we discuss the detectability of the signatures of CNO. 
The neutrino flux at Earth, $f_\nu^{({\rm f})}$, can be estimated by Eqs.~(23) and (24) of Ref.~\cite{Wu2015EffectsModel}.
In these equations, the effect of MSW is included and that of Earth effect is not included. The rotated frame of $e-x-y$ is also used  \cite{Dasgupta2008CollectiveNeutrinos}.
For convenience, we summarize the relevant parts. 
For normal mass hierarchy, the equations are
\begin{align}
    f_{\nu_e}^{({\rm f})} &= 
               s^2_{13} f_{\nu_e}^{({\rm a})}
    + c^2_{12} c^2_{13} f_{\nu_x}^{({\rm a})}
    + s^2_{12} c^2_{13} f_{\nu_y}^{({\rm a})}, \label{eq:nenmh}\\
f_{\bar{\nu}_e}^{({\rm f})} &=
      c^2_{12}c^2_{13} f_{\bar{\nu}_e}^{({\rm a})}
    + s^2_{12}c^2_{13} f_{\bar{\nu}_x}^{({\rm a})}
    +         s^2_{13} f_{\bar{\nu}_y}^{({\rm a})}.\label{eq:nbnmh}
\end{align}
For inverted mass hierarchy, the equations are
\begin{align}
    f_{\nu_e}^{({\rm f})} &= 
      s^2_{12} c^2_{13} f_{\nu_e}^{({\rm a})}
    + c^2_{12} c^2_{13} f_{\nu_x}^{({\rm a})}
    +          s^2_{13} f_{\nu_y}^{({\rm a})}, \label{eq:neimh}\\
    f_{\bar{\nu}_e}^{({\rm f})} &= 
               s^2_{13} f_{\bar{\nu}_e}^{({\rm a})}
    + s^2_{12} c^2_{13} f_{\bar{\nu}_x}^{({\rm a})}
    + c^2_{12} c^2_{13} f_{\bar{\nu}_y}^{({\rm a})}. \label{eq:nbimh}
\end{align}
In the above equations, $c_{ij}$ and $s_{ij}$ stand for $\cos\theta_{ij}$ and $\sin\theta_{ij}$, respectively.
We denote our spectrum after CNO as $f_{\nu}^{({\rm a})}$
\footnote{
In Ref.~\cite{Wu2015EffectsModel}, $f^{(a)}_{\nu}$ is denoted as $f^{(i)}_{\nu}$, where the superscript $(i)$ means the ``initial flux''. We change the superscript to $(a)$ in order to avoid confusion with the ``original flux'' before CNO. In our study, neutrinos that have not reached the point of high resonance is labeled as the initial flux that the authors define. See also Eq.~(19) for their definition of the variable \cite{Wu2015EffectsModel}.
}.

In the case of inverted mass hierarchy,
the anti-neutrino experience MSW resonances after CNO.
At the high resonance, $\bar{\nu}_e$ and $\bar{\nu}_y$ are completely swapped.
At the low resonance, $\bar{\nu}_e$ and $\bar{\nu}_x$ are mixed and 70\% of the $\bar{\nu}_e$ survives.
In short, we obtain an approximate equation from Eq.~\eqref{eq:nbimh}:
\begin{align}
    f_{\bar{\nu}_e}^{({\rm f})} &\sim
        0.7(1-\epsilon) f_{\bar{\nu}_e}^{({\rm o})} +
    (0.3 + 0.7\epsilon) f_{\bar{\nu}_X}^{({\rm o})}, \label{eq:nbimh-simple}
\end{align}
where the $f_{\nu}^{({\rm o})}$ represents the ``original flux'' and
$\epsilon$ is the survival probability of $\bar{\nu}_e$ just after CNO, i.e., 
$f_{\bar{\nu}_e}^{({\rm a})} = \epsilon f_{\bar{\nu}_e}^{({\rm o})} + (1-\epsilon)  f_{\bar{\nu}_X}^{({\rm o})}$.
Here conversion between $e-y$ is assumed. Namely we substituted the following equations to Eq.~\eqref{eq:nbimh},
$f_{\bar{\nu}_y}^{({\rm a})} = (1-\epsilon) f_{\bar{\nu}_e}^{({\rm o})} + \epsilon  f_{\bar{\nu}_X}^{({\rm o})}$
and $f_{\bar{\nu}_x}^{({\rm a})} = f_{\bar{\nu}_X}^{({\rm o})}$.
This assumption is most valid for the time between 100 ms to 300 ms post bounce.
Though $e-x$ and $e-y$ conversions can be seen during this phase (e.g., FIG.~\ref{fig:r-suv}),
the $e-x$ effect is not so prominent in intermediate and high energies relevant for detection. 

Roughly speaking, the survival probability of $\bar{\nu}_e$ after the occurrence of CNO is $0.3$ for $E>15\ {\rm MeV}$ (see FIG.~\ref{fig:T-E_SUV_G}).
Substituting $\epsilon =0.3$ into the equations above, we obtain
$f_{\bar{\nu}_e}^{({\rm a})} = 0.3 f_{\bar{\nu}_e}^{({\rm o})} + 0.7 f_{\bar{\nu}_X}^{({\rm o})}$.
This can be confirmed in the top panel of FIG.~\ref{fig:E-Fdetail}.
Above $\sim 15$ MeV, the red line, $f_{\bar{\nu}_e}^{({\rm a})}$, is closer to the blue line, $f_{\bar{\nu}_X}^{({\rm o})}$.
From Eq.~\eqref{eq:nbimh-simple}, we obtain
$ f_{\bar{\nu}_e}^{({\rm f})} = 0.49 f_{\bar{\nu}_e}^{({\rm o})} + 0.51 f_{\bar{\nu}_X}^{({\rm o})} $ for $\epsilon =0.3$.
This can be confirmed in the bottom panel of FIG.~\ref{fig:E-Fdetail}, where the red line, $f_{\bar{\nu}_e}^{({\rm f})}$, sits almost at midpoint between the blue line, $f_{\bar{\nu}_X}^{({\rm o})}$,
and the green line, $f_{\bar{\nu}_e}^{({\rm o})}$, for this energy range.

\subsubsection{Detection Property of $\bar{\nu}_e$}
The neutrinos that reach Earth can be detected by neutrino observation facilities. 
The main interaction for $\bar{\nu}_e$ is the inverse-beta decay.
The event  rate of the inverse-beta decay, $\frac{{\rm d} N}{{\rm d} t}\ [{\rm 1/s}]$, can be evaluated by the following equation \footnote{Here we ignore the dependence of kinetic energy of the scattered particle in the cross section. In general, we have to take into account that. For example, in the case of the scattering of neutrino and electron,
the kinetic energy of the electron should be considered in the equation. In the case of inverse-beta decay, the kinetic energy is identically determined and we do not have to include it explicitly in the equation.}:
\begin{equation}
    \frac{{\rm d} N}{{\rm d} t} = N_{\rm tar} \int_{E_{\rm th}}
    F \sigma  {\rm d}E, \label{eq:EventRate}
\end{equation}
where $N_{\rm tar}$ is the number of the target in the detector, $E_{\rm th}$ is the threshold energy of the detector, $F\ [{\rm /MeV/cm^2/s}]$ is the number flux of neutrino at earth
and $\sigma\ [{\rm cm^2}]$ is the cross section of the target to neutrinos.
The variables in the integral depend on the energy of the neutrino, $E\ [{\rm MeV}]$.
It should be noted that $F$ is proportional to the 
inverse square of the source distance.
The fluxes shown in FIG.~\ref{fig:E-Fdetail}
assume a source distance of 10 kpc, corresponding approximately to the distance to the Galactic center. 

For the case of HK detector,
we adopt
\begin{align}
    N_{\rm tar} &= N_{\rm A} \left(\frac{2M_{\rm H}}{M_{\rm H_2O}}\right) \rho_{\rm H_2O} V,
\end{align}
where $V$ is the volume of the detector, set to 220 kton, 
$N_{\rm A}$ is  Avogadro constant, and $\rho_{\rm H_2O}$ is the density of water. 
In the equation, 
$\left(\frac{2M_{\rm H}}{M_{\rm H_2O}}\right)$ is the mass fraction of hydrogen in ${\rm H_2O}$ and equals $\frac{2}{18}$.
We use the cross section of $\sigma = 9.77 \times 10^{-44} \left(\frac{E}{1\ [{\rm MeV}]}\right)^2\ [{\rm cm^2}]$. Including corrections of order $1/M_p$ to the cross section and kinematics \cite{Vogel1999Angular/m} yields typically 10--20\% reduction in event rates depending on detection threshold.
The threshold energy is set to $E_{\rm th} = 8.3 \, {\rm MeV}$ \cite{Kato2017NeutrinoDetections}.

\begin{figure}[htbp]
\includegraphics[width=0.95\linewidth]{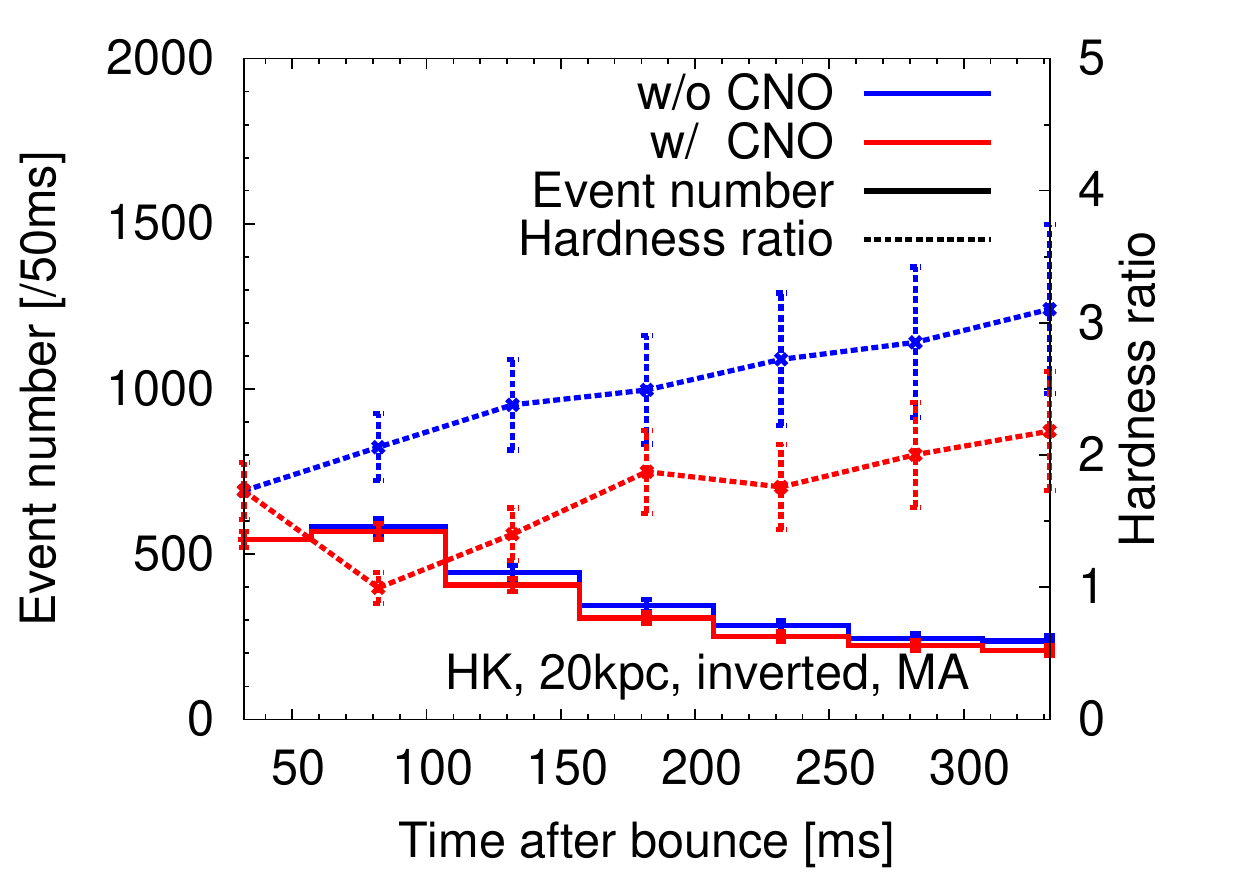}\\
\includegraphics[width=0.95\linewidth]{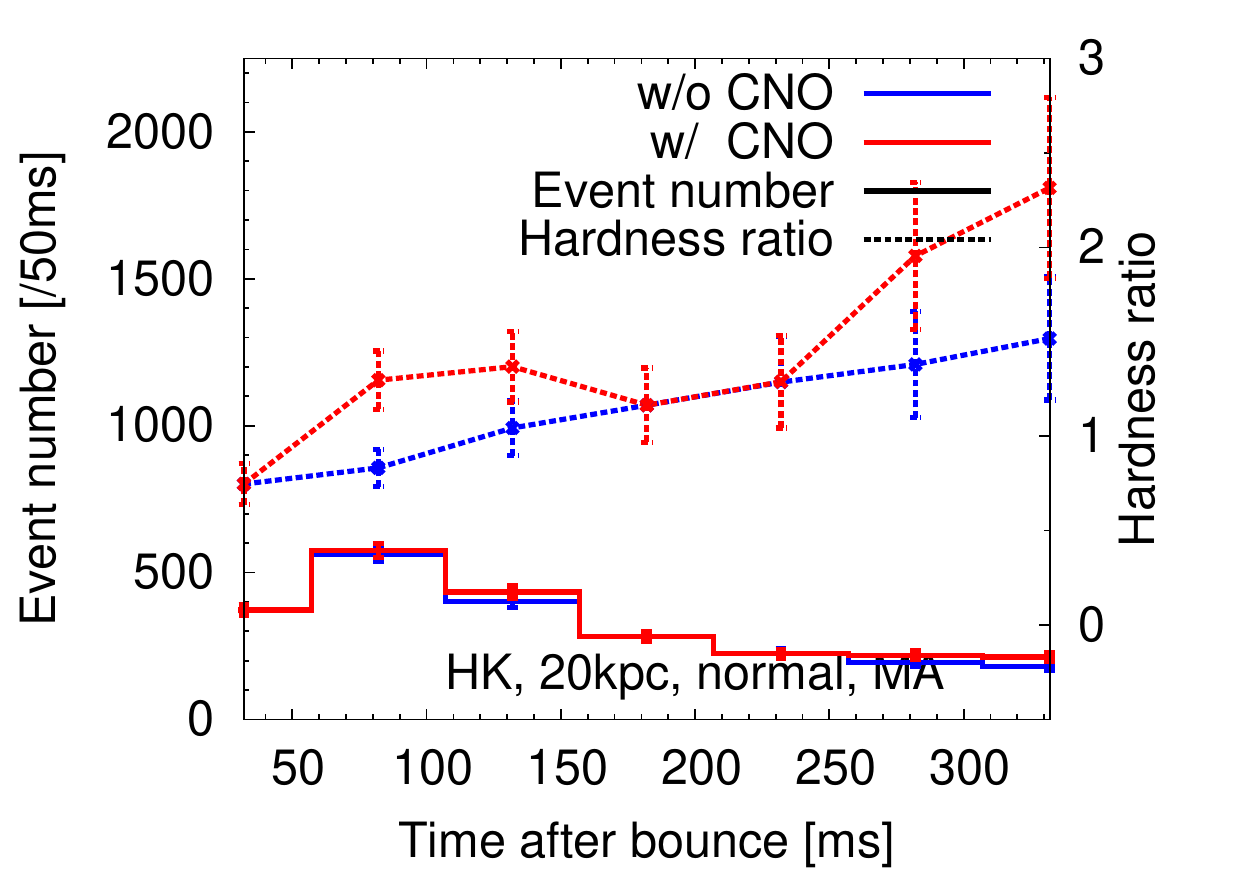}\\
\includegraphics[width=0.95\linewidth]{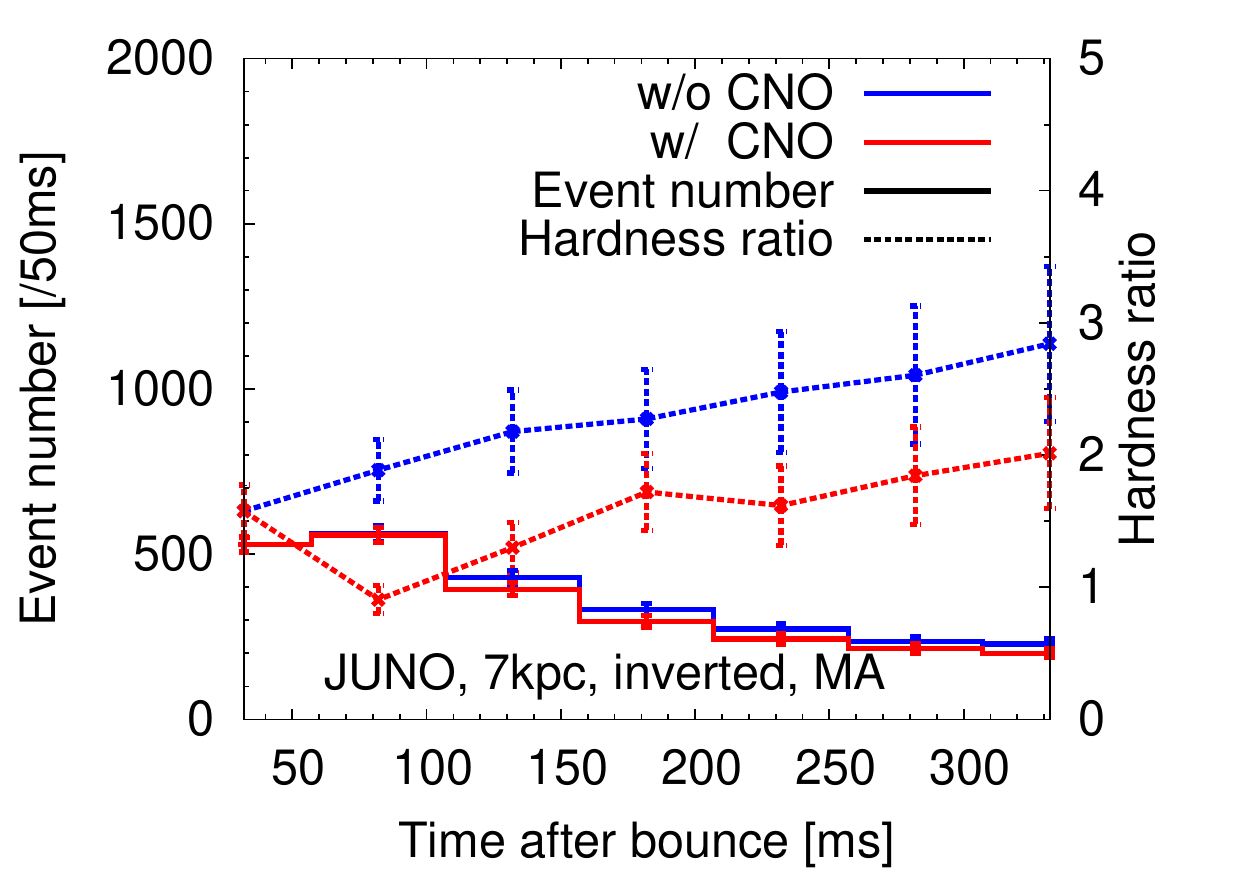}
\caption{
Evolution of event number in 50 ms bins (solid lines, left axis) and the hardness ratio (dotted lines, right axis).
Top: That of HK in the inverted mass hierarchy. A volume of 220 kton is adopted.
Middle: That of HK in the normal mass hierarchy. 
Bottom: That of JUNO in the inverted mass hierarchy. See the text for adopted detector parameters.
}
\label{fig:t-N_HK_KL}
\end{figure}

Thousand of neutrinos will be detected in every 50 ms bin if a supernova occurs near the Galactic center.
The event number in 50 ms bins, $N$, is shown in the top panel of FIG.~\ref{fig:t-N_HK_KL}
in solid lines (the left axis).
The definition of the number is $N = \frac{{\rm d} N}{{\rm d} t} \times \Delta t$ and  $\Delta t = 0.05\ [{\rm s}]$. 
The error bar of the line, $\pm \delta N$, is evaluated by Poisson error, i.e.,
$\delta N/N = 1/\sqrt{N}$.
The red/blue color represents the event number with/without CNO.
Naively, one may expect that the with/without CNO scenarios can be distinguished
since their difference  is larger than the Poisson error. 
However, this neglects other sources of errors coming from our limited knowledge on the progenitor.
For example, it is hard to know the detailed structure of the stellar core in reality, which 
strongly affects the neutrino luminosity \cite{OConnor2013TheSupernovae,Nakazato2013SupernovaCooling}.
And if the explosion happen in our Galaxy, the evaluation of its distance is oftentimes difficult.
The distance to the supernova can easily change the neutrino luminosity \cite{Horiuchi2017EstimatingNeutrinos}.
These uncertainties can be larger than the difference between with/without CNO.

While it may be difficult to distinguish the with/without CNO scenarios based solely on the event number,
there are ways to circumvent much of the additional systematic uncertainties. 
To see the effect of CNO, 
we define the hardness ratio, $ R_{\rm H/L}$, following Ref.~\cite{Kawagoe2009NeutrinoExplosions}:
\begin{equation}
    R_{\rm H/L} = \frac{N_{E_c < E}}{N_{E < E_c}}, \label{eq:ratio}
\end{equation}
where $N_{E_c < E}$ and $N_{E < E_c}$ are event numbers whose neutrino energy is above $E_c$ and below $E_c$, respectively.
The error of the ratio is given by the following equation:
\begin{align}
    \delta R_{\rm H/L}/R_{\rm H/L} 
    &= \frac{\delta N_{E_c < E}}{N_{E_c < E}} + \frac{\delta N_{E < E_c}}{N_{E < E_c}}\nonumber\\
    &= \frac{1}{\sqrt{N_{E_c < E}}} + \frac{1}{\sqrt{N_{E < E_c}}}.
\end{align}
This ratio is sensitive to the neutrino average energy and not 
sensitive to the integrated flux. That means the error from 
the stellar structure and distance of the source does not 
strongly affect the ratio. 

The evolution of the hardness ratio with $E_c= 20\ {\rm MeV}$ is
shown in  the top panel of FIG.~\ref{fig:t-N_HK_KL}
in dotted lines (the right axis).
The blue line corresponds the ratio without CNO.
Due to MSW high-resonance,
the spectrum of $\bar{\nu}_e$ at Earth is exactly that of the original $\nu_X$.
The red line is depicted with the effect of CNO.
CNO changes the spectrum and some fraction of 
the original spectrum of $\bar{\nu}_e$ remains in the 
spectrum at Earth. Compare to the case without CNO,
the spectrum with CNO becomes softer since the average energy of 
the original $\bar{\nu}_e$ is lower than that of the original $\nu_X$.

When CNO happens, the hardness ratio suddenly becomes smaller.
This feature is easy to distinguish from that of without CNO since the 
ratio naturally tend to increase as time goes by.
The latter trend is seen in the blue dotted line in the top panel of FIG.~\ref{fig:t-N_HK_KL}.
During this phase,
the neutron star is shrinking. 
Then the neutrino spectrum naturally evolves to become hard as 
the neutrinosphere becomes smaller and the 
effective temperature becomes higher. 
Since the effect of CNO is the opposite of this generic trend, 
it can be easily identified. 
The error bar of the hardness ratio at 20 kpc source is less than the difference between the models with and without CNO.
 We can distinguish the two models even if we take the $1\sigma$ Poisson error into account.

\begin{figure}[t]
\includegraphics[width=0.95\linewidth]{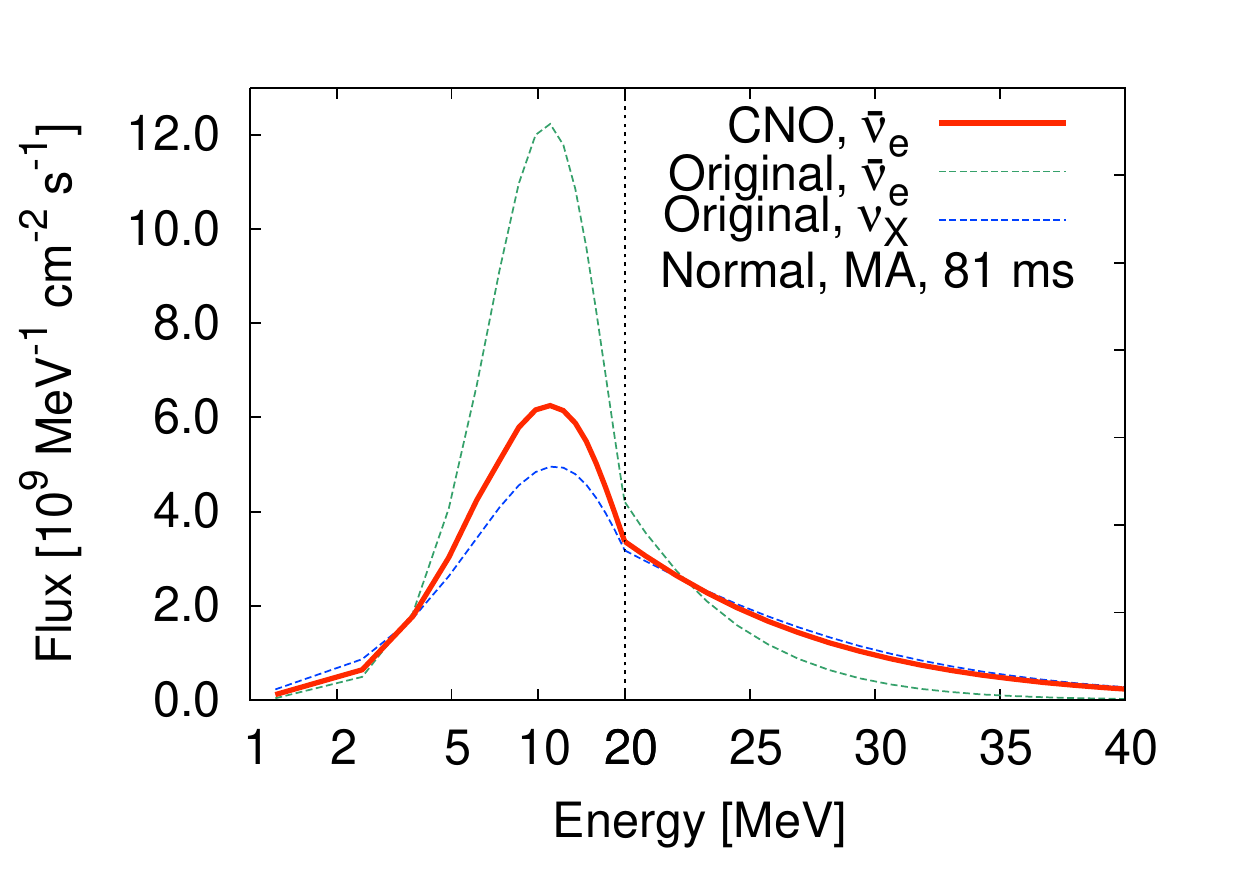}\\
\includegraphics[width=0.95\linewidth]{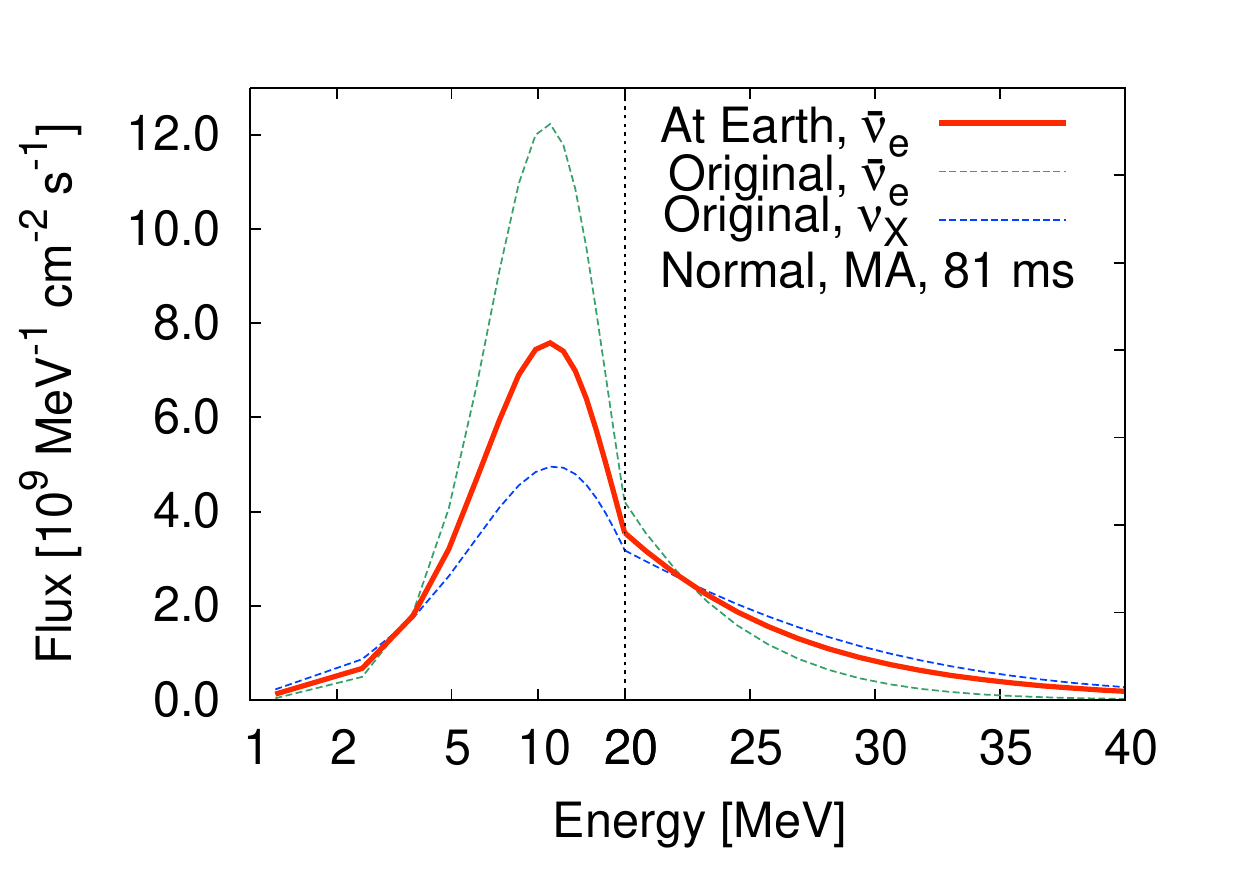}
\caption{
Same to FIG.~\ref{fig:E-Fdetail} but for the normal mass hierarchy.
The time snapshot of 81 ms is used.
}
\label{fig:E-Fdetail_normal}
\end{figure}

In the case of normal mass hierarchy,
the hardness ratio of $\bar{\nu}_e$ becomes hard with the effect of CNO.
We calculate the flux at Earth using Eq.~\eqref{eq:nbnmh}.
An approximate equation of Eq.~\eqref{eq:nbnmh} is 
\begin{align}
    f_{\bar{\nu}_e}^{({\rm f})} &\sim (0.3+0.4\epsilon) f_{\bar{\nu}_e}^{({\rm o})} +(0.7 - 0.4\epsilon)f_{\bar{\nu}_X}^{({\rm o})}, \label{eq:nbimh-app}
\end{align}
where $\epsilon$ is the survival probability of $\bar{\nu}_e$ just after CNO (conversion between $e-x$ is assumed).
The event rate and the hardness ratio with $E_c=20\ {\rm MeV}$ are given in 
the middle panel of FIG.~\ref{fig:t-N_HK_KL}.
The blue line corresponds that of without CNO model, i.e., $\epsilon = 1$.
In the normal mass hierarchy, MSW high resonance 
does not affect the spectrum of $\bar{\nu}_e$
and 70\% of  $\bar{\nu}_e$ survives at MSW low resonance (see Eq.~\eqref{eq:nbimh-app}).
As a result, the spectrum at Earth is similar to that of 
the original $\bar{\nu}_e$ and the hardness ratio is low.
The red lines show the results with CNO.
In this case, CNO decreases the survival probability of $\bar{\nu}_e$.
The spectrum of $\bar{\nu}_e$ at Earth contains a 
large fraction from the original $\nu_X$.
As a result, the hardness ratio becomes higher.

The value of $\epsilon$ can be seen in the bottom panel FIG.~\ref{fig:T-E_SUV_G}.
The spectra at 81 ms post bounce is shown in FIG.~\ref{fig:E-Fdetail_normal}.
In this time, the survival probability, $\epsilon$, is $0.2$ and a large fraction of the electron anti-neutrinos is converted to $x$ anti-neutrinos. The top panel of FIG.~\ref{fig:E-Fdetail_normal} shows the spectra after CNO (1500 km).
The spectrum of electron anti-neutrino is closer to that of the original $\bar{\nu}_X$.
After CNO, the neutrinos experience MSW-low resonance and the $x$ anti-neutrinos  and electron anti-neutrinos are mixed.
In this case, the $x$ anti-neutrino contains a large fraction of original $\bar{\nu}_e$ and the 
electron anti-neutrino spectrum after MSW-low resonance becomes softer.
This is shown in the bottom panel of FIG.~\ref{fig:E-Fdetail_normal}.
The electron anti-neutrino spectrum (red curve) is softer than that of that after CNO (red curve in the top panel of FIG.~\ref{fig:E-Fdetail_normal}). 

At the onset of CNO, we will therefore see a rapid rise of the hardness ratio.
Unfortunately, this leaves from level of degeneracy with 
the natural rise of the hardness due to the evolution of the proto neutrino star.
One would have to compare the rise times of the hardness ratio 
to draw a robust conclusion.
While the error bar of the hardness ratio at a source distance of 20 kpc is smaller than the difference of the models with and without CNO, 
we have to consider the degeneracy. We conclude that the source distance should be less than $\sim 10$ kpc to test the presence of CNO.
The statistical error will improve linearly with distance, increasing the potential to distinguish between models. A statistical statement will require us to marginalize over the degeneracy stated above and also work with multiple time bins, which are beyond the scope of this work, but, e.g., a $3\sigma$ statement should require distances of less than 3.3 kpc.

The value of the hardness ratio depends on the detector.
The event rate and the hardness ratio of JUNO are given in 
the bottom panel of FIG.~\ref{fig:t-N_HK_KL}.
Since the energy threshold of JUNO is lower than HK,
JUNO can capture low energy neutrinos.
Reflecting that feature, the value of the hardness ratio becomes low
compared to that of HK.
However, the overall features are not so different.
Due to the smaller volume of JUNO, the statistical errors are comparable to the difference between the with and without CNO predictions, showing the source distance 
should be less than some 5 kpc to distinguish the effect of CNO.
We use Eq.~\eqref{eq:EventRate} to evaluate the event number.
We assume that JUNO is a 20 kton detector \cite{An2016NeutrinoJUNO}.
First we evaluate the event rate in KamLAND and later multiply a factor coming from the volume difference of 
$(20/0.7)$ to obtain the rate in JUNO.
The number of target proton in KamLAND, 
$ N_{\rm tar}$, is $5.98\times 10^{31}$ for each 0.7 kton, fiducial volume \cite{Gando2016AKAMLAND}.
It should be noted that KamLAND uses dodecane and the density and mass ratio for ${\rm H_2O}$ cannot be applied.
The cross section is same as that of HK. The energy threshold is 1.8 MeV.

\subsubsection{Detection Property of $\nu_e$}
CNO is also expected in the electron neutrinos, where the oscillation property is as interesting as that of 
electron anti-neutrinos. Although it is difficult to detect larger numbers of clean $\nu_e$s with detectors currently in operation, the future large-volume liquid argon detector, DUNE, is expected to change this. To prepare for this era, we next predict the oscillation property in $\nu_e$.

\begin{figure}[t]
\includegraphics[width=0.95\linewidth]{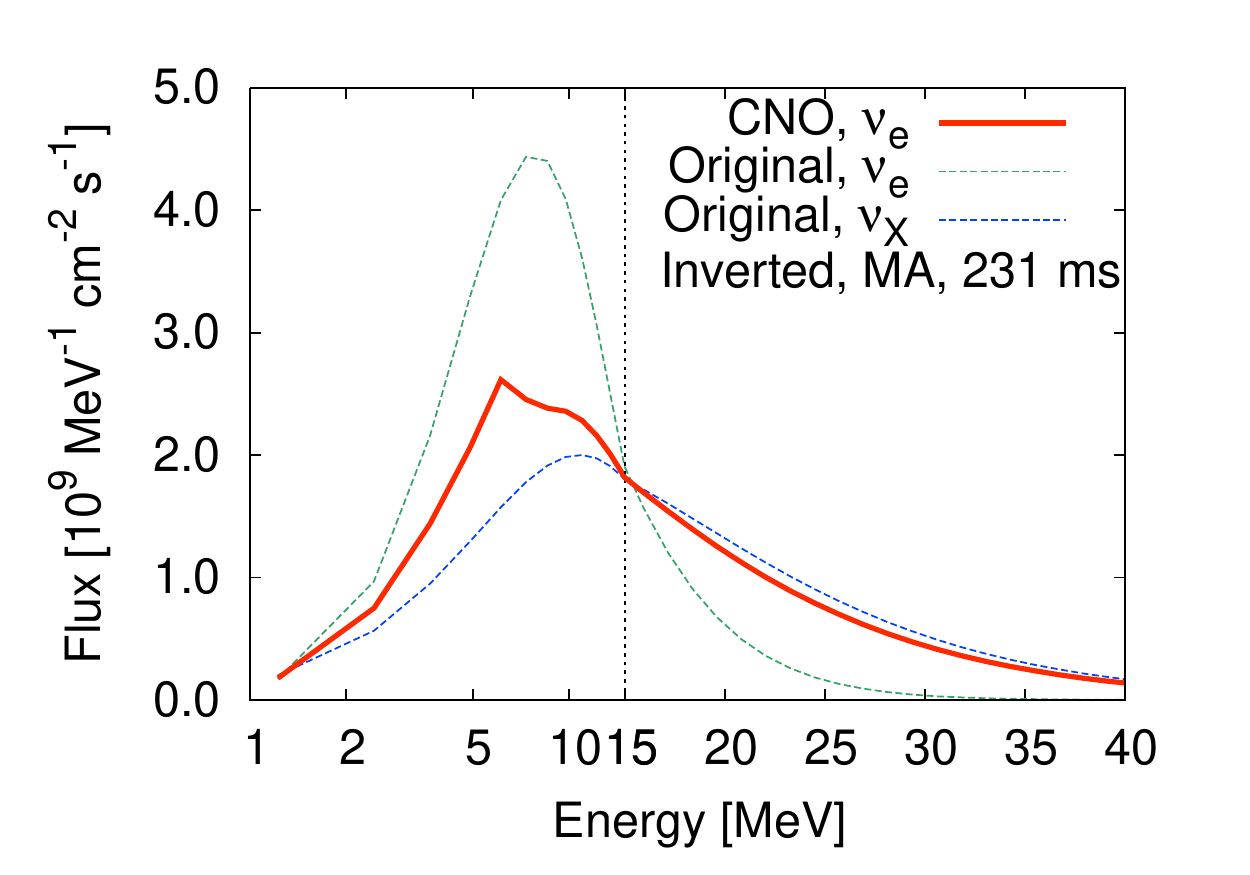}\\
\includegraphics[width=0.95\linewidth]{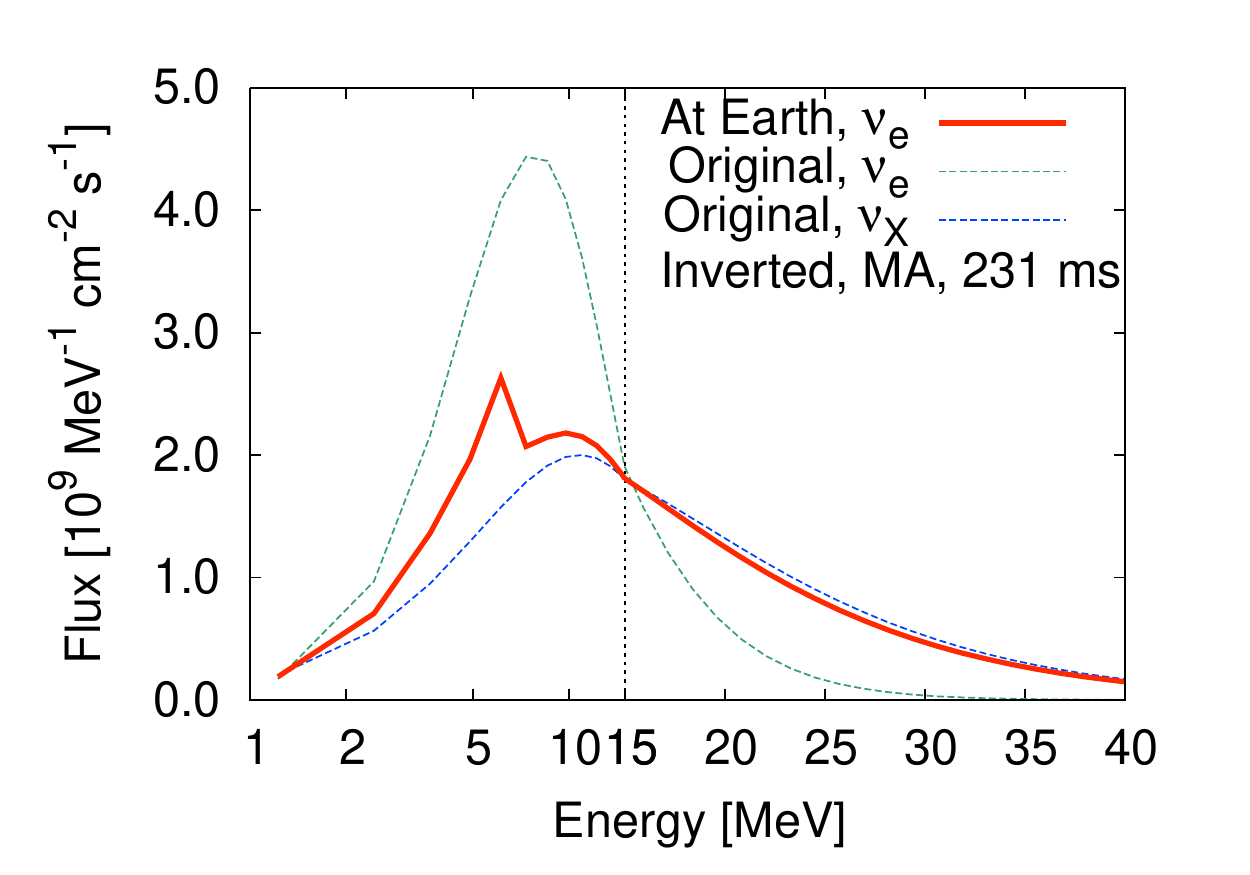}
\caption{
Same as FIG.~\ref{fig:E-Fdetail}, but for electron neutrinos.
Note that in both panels, the horizontal axis is logarithmic for $E<15\ {\rm MeV}$ and liner for $E>15\ {\rm MeV}$ in the both panels (indicated by the vertical dashed line).
}
\label{fig:E-Fdetailnue}
\end{figure}

The top panel of FIG.~\ref{fig:E-Fdetailnue} shows the spectrum
in the case of the inverted mass hierarchy.
After CNO, the neutrinos experience the MSW low resonance, and 
30\% of the $\nu_e$ survives. 
From Eq.~\eqref{eq:neimh},
\begin{align}
    f_{\nu_e}^{({\rm f})} &\sim 0.3\epsilon f_{\nu_e}^{({\rm o})}+(1.0 - 0.3\epsilon)f_{\nu_X}^{({\rm o})}, \label{eq:neimh-app}
\end{align}
where $\epsilon$ is the survival probability just after CNO ($e-y$ conversion is assumed).
MSW low resonance significantly makes the fraction of the original $\nu_e$ at Earth lower:
if CNO does not happen, the fraction is maximum at 30\% and CNO makes the fraction lower.
The bottom panel of FIG.~\ref{fig:E-Fdetailnue} shows the spectrum at Earth.
The spectrum (red curve) almost looks like that of original $\nu_X$ (blue curve).

For normal mass hierarchy,
we obtain an approximate formula from Eq.~\eqref{eq:nenmh} by similar analysis as above:  
\begin{align}
    f_{\nu_e}^{({\rm f})} &\sim (0.7-0.7\epsilon) f_{\nu_e}^{({\rm o})} +(0.3 + 0.7\epsilon)f_{\nu_X}^{({\rm o})}, \label{eq:nenmh-app}
\end{align}
where $\epsilon$ is the survival probability just after CNO ($e-x$ conversion is assumed).

\begin{figure}[t]
\includegraphics[width=0.95\linewidth]{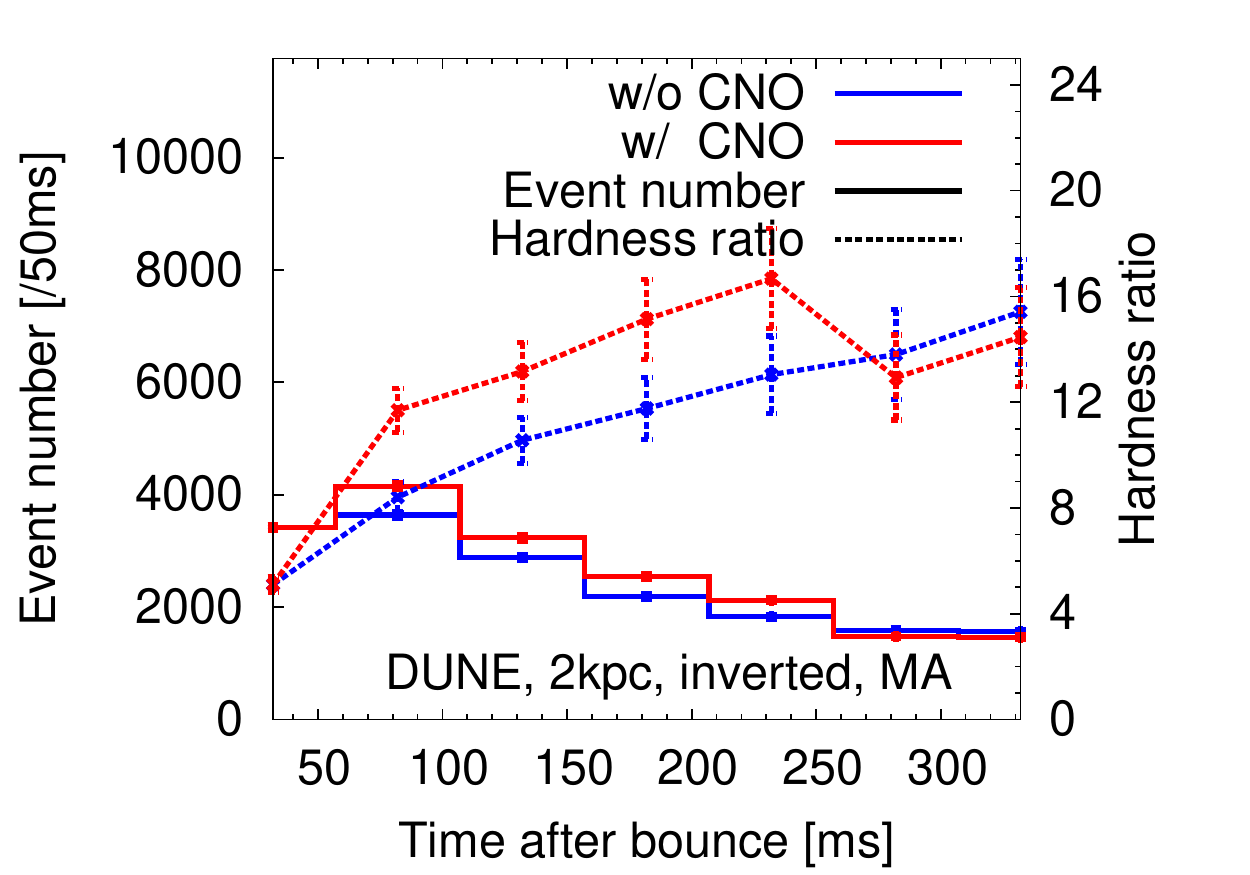}\\
\includegraphics[width=0.95\linewidth]{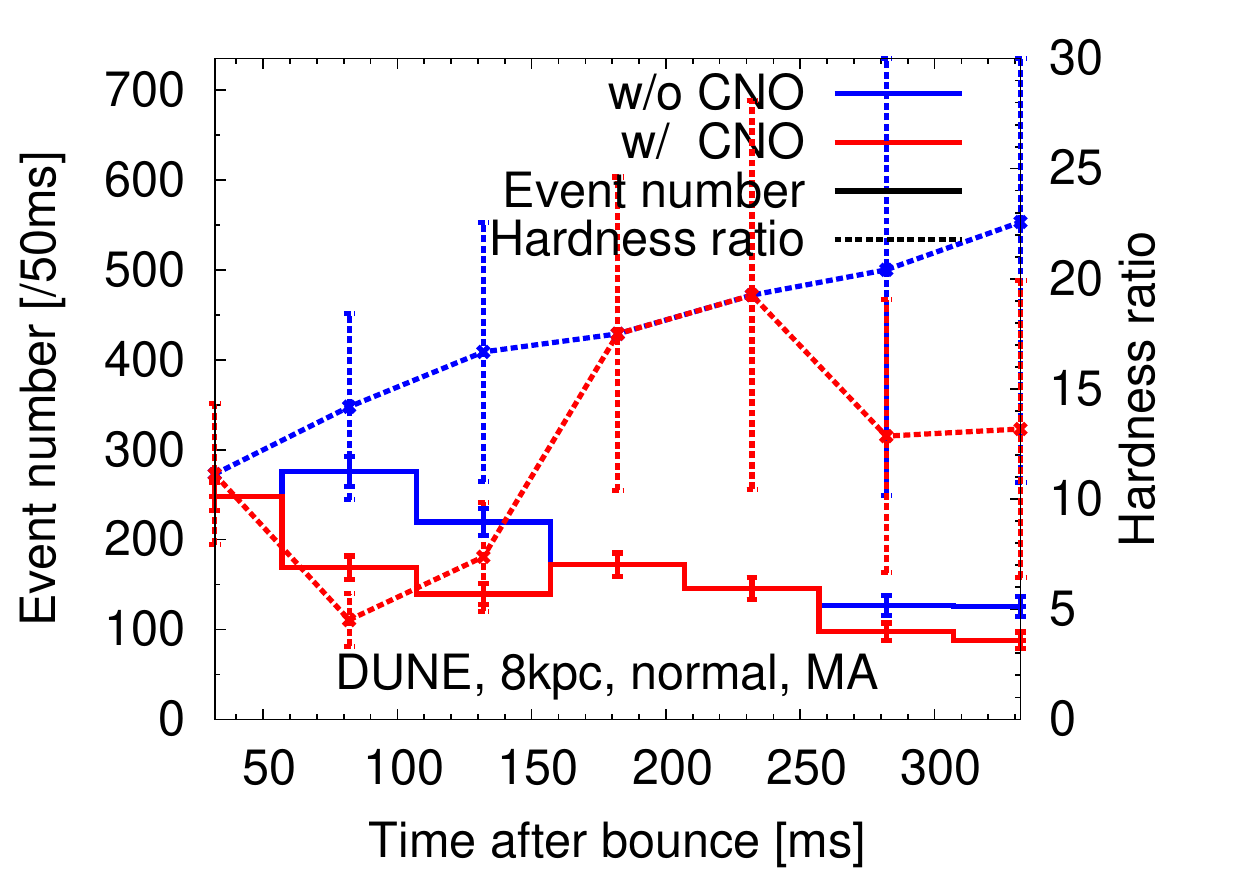}
\caption{
Event number and hardness ratio in 50 ms bins, both for $\nu_e$ detected by DUNE. Note that a lower $E_c$ is adopted that in FIG.~\ref{fig:t-N_HK_KL}.
Top: the case of inverted mass hierarchy and a source distance of 2 kpc.
Bottom: the case of normal mass hierarchy and a source distance of 8 kpc.
}
\label{fig:t-NR_DUNE}
\end{figure}

The $\nu_e$ emitted in the supernova can be detected in large numbers.
Of the existing and planned neutrino detectors, DUNE is the primary detector 
with an expected high-statistics, clean $\nu_e$ signal \cite{Scholberg:2012id}. 
We use the cross section of the primary charge-current interaction on liquid
argon, $\nu_e+{^{40}}{\rm Ar}\rightarrow e^{-} + {^{40}}{\rm K}^{*}$ based on 
the random phase approximation scheme of Ref.~\cite{Kolbe:2003}.
Using Eq.~\eqref{eq:EventRate} we compute the event rates due to this reaction. We evaluate the number of the target nuclei taking DUNE's total fiducial volume to be 40 kton.
We adopt a detection threshold of 5 MeV $\nu_e$ energy and for simplicity
assume a detection efficiency of 100\%. The true threshold and efficiency remain to be determined. In reality, the supernova neutrino's low
energy means that the interaction products may only leave stub-like tracks and blips
in the liquid argon time-projection chamber; also, the signal may be vulnerable to
radioactive and cosmogenic backgrounds \cite{Ankowski:2016lab}. More work is ongoing 
to understand the efficiency as a function of detector configuration. 
We estimate the number of events in 50 ms bins and the hardness ratio with 
$E_c=15\ {\rm MeV}$, and show their time evolutions in FIG.~\ref{fig:t-NR_DUNE}.
 
In the case of inverted mass hierarchy,
the spectrum of $\nu_e$ becomes hard after CNO. 
In the top panel of FIG.~\ref{fig:t-NR_DUNE},
the blue (red) line is the event number and the hardness ratio without (with) CNO.
Equation \eqref{eq:neimh-app}
shows that the CNO decreases the fraction of original $\nu_e$ in the spectrum at Earth and 
increases the fraction of original $\nu_X$.
As a result, the spectrum becomes hard after CNO happen.
That feature appears before 281 ms.
In the later explosion phase, such as $331$ and $381$ ms, however, $x-y$ mixing after $e-y$ conversion as shown in bottom panel of FIG.~\ref{fig:r-suv} occurs actively in the neutrino sector. Such prominent three flavor mixing makes the hardness ratio of $\nu_{e}$ small in the inverted mass hierarchy. This behavior can be explained  by considering the survival probability $\eta$ in the $x-y$ conversions. The final $\nu_{e}$ flux on Earth is described by
\begin{align}
\begin{split}
    f_{\nu_e}^{({\rm f})} &\sim [0.3\epsilon+0.7(1-\eta)(1-\epsilon)] f_{\nu_e}^{({\rm o})}\\
    &+[1.0 - 0.3\epsilon-0.7(1-\eta)(1-\epsilon)]f_{\nu_X}^{({\rm o})}, \label{eq:neimh-app-2}
    \end{split}
\end{align}
which reproduces Eq.(\ref{eq:neimh-app})  under the two flavor limit in $e-y$ sector  ($\eta\to1$). As shown in Eq.(\ref{eq:neimh-app-2}), three flavor mixing  ($0\leq\eta<1$) increases the fraction of original $\nu_{e}$ which prevents hard $\nu_{e}$ spectrum at Earth. 
To make the error bar of the hardness ratio smaller than the model difference, the source distance should be less than $\sim$ 2 kpc.

On the other hand,
CNO makes the spectrum soft in the case of normal mass hierarchy regardless of three flavor mixing.
The bottom panel of FIG.~\ref{fig:t-NR_DUNE} shows the time evolution of event rate and the hardness ratio for the normal mass hierarchy. 
The blue line is without CNO, i.e., $\epsilon=1$ in Eq.~\eqref{eq:nenmh-app}.
Here, the survival probability of $\nu_e$ at Earth is $0$ by the MSW high resonance.
CNO cancels the effect of the MSW high resonance.
Then the survival probability become finite and at most 70\% (see  Eq.~\eqref{eq:nenmh-app} and substitute $\epsilon = 0$).
Although the MSW low resonance also decreases the survival probability,
still a fraction of the original $\nu_e$ remains at Earth.
This makes the spectrum soft.
Since the amplitude of this softening is large compared to the error bar of the hardness ratio, 
the statistical errors remain smaller than the model difference out to source distance of $\sim 8\ {\rm kpc}$.

Compared to the case in HK, we need closer source distance to distinguish the effect of CNO in DUNE.
The larger statistical error originates from the smaller $N_{E<E_c}$.
Since we do not optimise the value of $E_c$ here,
changing the value of $E_c$ could decrease
the error in the future studies.

\subsubsection{Synergistic observation}
We summarize the effect of CNO in Table \ref{tab:summary}.
In the case of the inverted mass hierarchy, the 
$\bar{\nu}_e$ spectrum without CNO is hard since the original-$\nu_X$ spectrum is observed at Earth (see the second column of the third row).
The spectrum becomes soft if CNO takes place (see the third column of the third row).
On the other hand, the spectrum of $\nu_e$ is soft before the occurrence of CNO (see Eq.~\eqref{eq:neimh-app}). 
Here, CNO makes the spectrum hard before the three flavor mixing in CNO is switched on.
After the non-linear three flavor mixing, the fraction of original $\nu_e$ increases and the spectrum becomes soft  (see third column of the fourth row). To warn of this complicated behavior, we add a $^*$ mark in the table.

The role of CNO in the normal mass hierarchy has the opposite effect as in the inverted mass hierarchy.
Namely, the soft spectrum of $\bar{\nu}_e$ becomes harder with the onset of CNO.
The CNO make the hard spectrum of $\nu_e$ softer. These are summarized in the last two columns of table \ref{tab:summary}.

Interestingly, the effect of CNO is such that when the spectrum of $\bar{\nu}_e$ is harder, that of $\nu_e$ is softer. 
This means synergistic observations of $\bar{\nu}_e$ and $\nu_e$ would be valuable to look for the occurrence of CNO. In this respect, the result from HK and DUNE will complement reach other very strongly.
The horizon for joint observation appears to be slightly smaller than 10 kpc. For example, DUNE may capture the onset of CNO in the inverted mass hierarchy for sources closer than $\sim 2\ {\rm kpc}$ (see top panel of FIG.~\ref{fig:t-NR_DUNE}).

\begin{table}[htbp]
    \centering
    \begin{tabular}{c|cccc}
                Hierarchy & Inverted &  Inverted     & Normal & Normal    \\
                CNO       & Off      & On            & Off    & On \\
                \hline
   $\bar{\nu}_e$ spectrum & Hard     & Soft          & Soft   & Hard \\
       ${\nu}_e$ spectrum & Soft     & Hard$^*$      & Hard   & Soft \\
    \end{tabular}
    \caption{Summary of the effect of CNO. See text for the meaning of $*$.}
    \label{tab:summary}
\end{table}

\section{Summary and Discussion}\label{sec:summary}

We performed neutrino radiation hydrodynamic simulation of a 8.8 \Msun progenitor and 
used the profile of the simulation to investigate the impact of collective neutrino oscillation (CNO)
on the detected neutrinos on Earth. We considered  Hyper-Kamiokande (HK), JUNO, and DUNE, 
and evaluated detectability defining a hardness ratio of the observed neutrino spectra. 
Our findings are summarized as follows.
\begin{itemize}
    \item CNO happens after 100 ms after bounce; it is suppressed by matter induced decoherence before this time. The dilute envelope structure of the progenitor is the main reason for such an early emergence of CNO. The precise time can be affected by the spectral shape.
    \item In the case of inverted mass hierarchy, the spectrum of $\bar{\nu}_e$ becomes softer after CNO sets in. The hardness ratio that we defined in Eq.~(\ref{eq:ratio}) is helpful to identify the onset of the CNO effect. HK can distinguish this effect when the supernova happens within a distance of $\sim 10$ kpc.  
    \item In the case of normal mass hierarchy, the spectrum of $\nu_e$ becomes softer after CNO happens. DUNE can distinguish this effect when the supernova happens within a distance of $\sim 10$ kpc. 
    \item If the spectrum of $\bar{\nu}_e$ becomes softer due to CNO, the spectrum of $\nu_e$ becomes harder (and vice versa). This provides a synergistic opportunity to combine the $\bar{\nu}_e$ and $\nu_e$ from HK and DUNE as a valuable method to test the occurrence of CNO.
\end{itemize}

There are several limitations in this study.
First, we finished our hydrodynamic simulations at 331 ms after bounce, since the 
density of the envelope becomes too low and protrudes the region of our tabulated EoS.
Due to that, we cannot investigate how long this CNO continues. 
Second, for the neutrino oscillation part, 
we have not included
the effect of the neutrino-nucleon interactions \cite{Shalgar2019}, 
the feed back of the oscillation to the hydrodynamics \cite{Stapleford2019}, 
the effect of the halo \cite{Cherry2013HaloBurst,Zaizen2019}, 
multi-azimuthal-angle instability \cite{Sawyer2009MultiangleSystems},
fast flavor conversion \cite{Abbar2019OnModels,Abbar2019,DelfanAzari2019LinearSimulations,Milad2019}
and non-standard neutrino self-interactions \cite{Stapleford2016NonstandardSupernovae,Dighe2018NonstandardConversions}, all of which can affect the resulting patterns. More studies will be needed to elucidate their effects and to draw robust conclusions about their detectability and differentiation.  

\acknowledgments
This study was supported in part by the Grants-in-Aid for the Scientific Research of Japan Society for the Promotion of Science (JSPS, Nos. 
JP17H01130, 
JP17K14306, 
JP18H01212, 
JP17K17655, 
JP19J13632), 
the Ministry of Education, Science and Culture of Japan (MEXT, Nos.
JP15H01039, 
JP17H06357, 
JP17H06364, 
JP17H05206, 
JP26104001, 
JP26104007, 
JP19H05803), 
JICFuS as a priority issue to be tackled by using Post `K' Computer.
This work is also supported by the NINS program for cross-disciplinary
study (Grant Numbers 01321802 and 01311904) on Turbulence, Transport,
and Heating Dynamics in Laboratory and Solar/Astrophysical Plasmas:
"SoLaBo-X”.
The work of SH is supported by the US Department of Energy Office of Science under award number DE-SC0020262, NSF Grant number AST-1908960, and NSF Grant number PHY-1914409.  

\bibliographystyle{h-physrev} 
\bibliography{ref,refByhand}

\appendix

\section{Dependence of position of neutrinosphere}\label{sec:neusphere}

In our simulations, we employ the single-bulb model and use a fixed neutrinosphere radius of $R=30$ km irrespective of the postbounce time \cite{Duan2006CollectiveSupernovae}. We should keep in mind that, originally, the neutrinosphere radius depends on the explosion time, as well as the neutrino energy and neutrino species. Further study is required to consider all of these effects on the CNO. Some previous works \cite{Sawyer:2015dsa,Chakraborty:2016lct,Abbar:2017pkh} employ a multi-bulb model which incorporates the flavor dependence in the neutrinosphere. Here, we show the impacts of the neutrinosphere radius, keeping to the simple bulb model but by changing the radius of the  neutrinosphere from $30$ km to $50$ km. Since the neutrinosphere radius continues to shrink with post bounce time, our simple demonstration is helpful to understand the uncertainty caused by the time-evolution of the neutrinosphere.\\

\begin{figure}[htbp]
\includegraphics[width=0.95\linewidth]{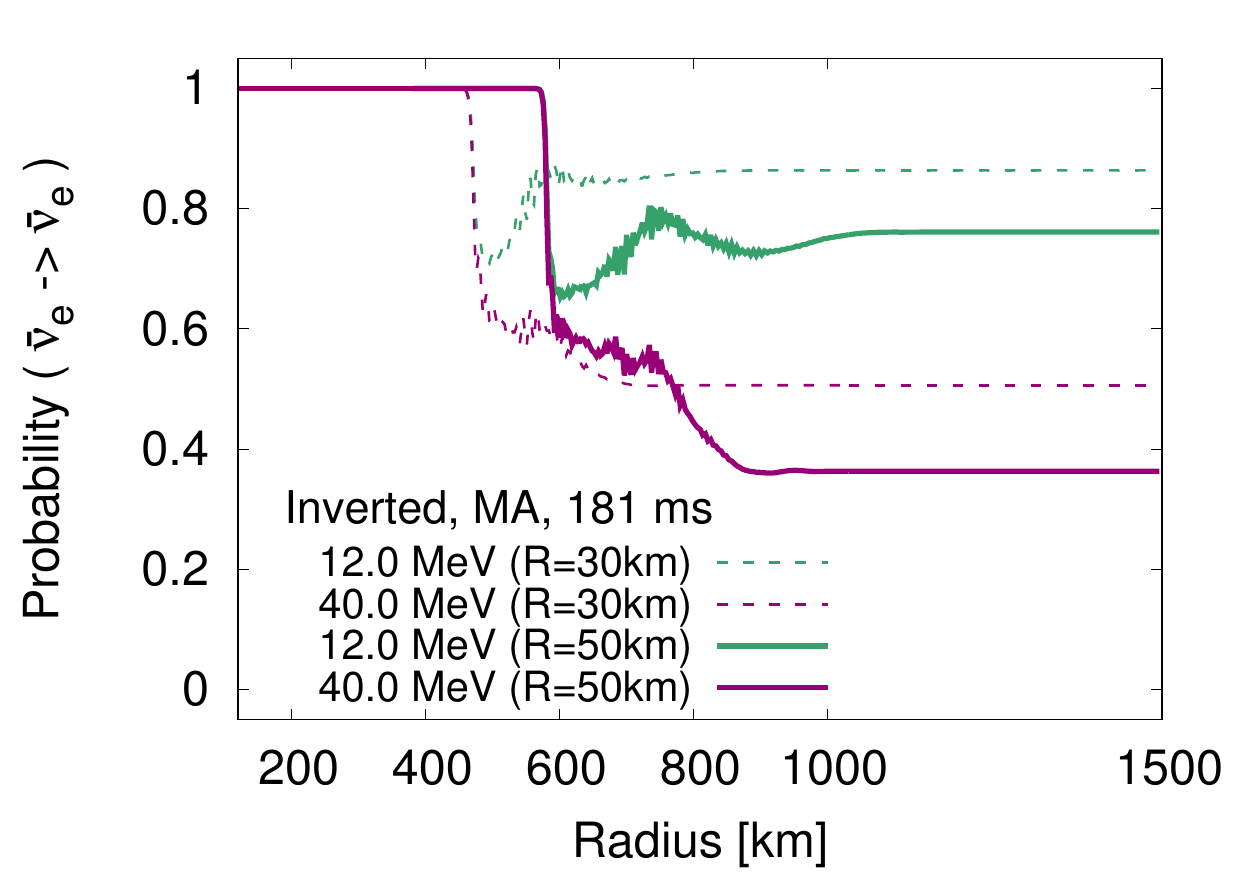}
\caption{
The comparison of conversion probabilities of $\bar{\nu}_{e}$ in the inverted mass hierarchy at $181\ {\rm ms}$ post bounce. The solid (dash) lines represent radial profiles of conversion probabilities in the bulb model assuming $R=30 (50)$ km, respectively. 
}
\label{fig:r-suv_232_inv}
\end{figure}

\begin{figure}[htbp]
\includegraphics[width=0.95\linewidth]{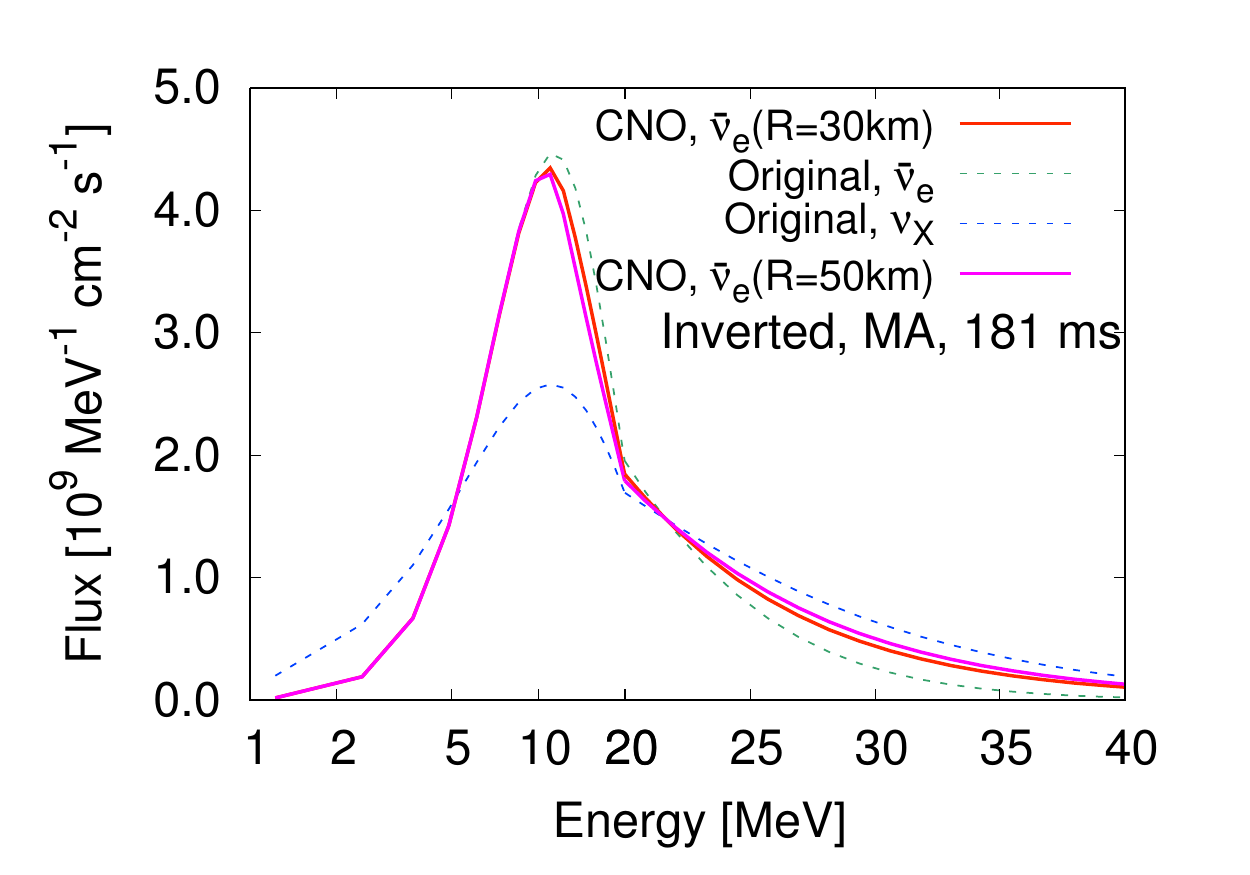}
\caption{
Comparison of $\bar{\nu}_{e}$ spectra after CNO. The red (magenda) line 
represents $\bar{\nu}_{e}$ spectra after CNO in $R=30$ ($50$) km model, respectively.
As FIG.~\ref{fig:E-Fdetail}, the flux is scaled to that for 10kpc source.
}
\label{fig:E-Fdetail_232_compare}
\end{figure}

FIG.~\ref{fig:r-suv_232_inv} shows the evolution of survival probabilities of $\bar{\nu}_{e}$ in the inverted mass hierarchy at $181$ ms. In this explosion epoch, CNO can be regarded as two flavor oscillations in the $e-y$ sector because of the decoupling of the $x$ type neutrinos. The onset of CNO is $455$ $(570)$ km in case of $R=30$ $(50)$ km, respectively. Such delayed non-linear effects in the large neutrinosphere radius is well explained by the relation between the neutrinosphere radius and instability radius: $r_{\rm{inst}}\propto R^{1/2}$ \cite{Duan:2010bf}. In multiangle simulations, significant non-linear flavor transitions occur when the dispersion of nonlinear self interactions are comparable with that of vacuum Hamiltonian \cite{Duan:2010bf}.

Final values of survival probabilities after CNO ($r>1000$ km) are also sensitive to the neutrinosphere radius. Flavor mixings in the larger neutrinosphere ($R=50$ km) are more prominent than in the smaller radius model ($R=30$ km). The strength of the matter potential is smaller in the outer region because of decreasing baryon density. The angular dispersion in the matter potential \cite{EstebanPretel:2008ni,Duan:2010bf} does not suppress the coherence of non-linear flavor transitions if the onset of flavor transitions is delayed. Therefore, CNO occurs more actively in the large neutrinosphere model. A spectral swap after CNO can be seen in FIG.~\ref{fig:E-Fdetail_232_compare}. The spectra of electron antineutrinos in both models are similar in low energy region ($E<E_{c}=20$ MeV). However, at high energies ($E>E_{c}$), the spectrum of $\bar{\nu}_{e}$ is slightly closer to the original spectrum of $\bar{\nu}_{X}$ in the $R=50$ km model compared with the $R=30$ km model. This enhanced spectral swap reflects active non-linear flavor transitions without suffering from strong multiangle matter suppression as shown in FIG.~\ref{fig:r-suv_232_inv}. Such increased energetic $\bar{\nu}_{e}$ in the $R=50$ km model makes the hardness ratio of $\bar{\nu}_{e}$ softer, which would be helpful to distinguish the effect of CNO. \\

So far, we have shown the result of CNO at $181$ ms by imposing $R=50$ km, but properties of delayed CNO as confirmed in FIGs.~\ref{fig:r-suv_232_inv},\ref{fig:E-Fdetail_232_compare} are also common in other explosion phases. The neutrinosphere radius becomes smaller as the explosion proceeds. Therefore, it seems that the hardness ratio of $\bar{\nu}_{e}$ as shown in FIG.~\ref{fig:t-N_HK_KL} tends to be softer (harder) in earlier (later) explosion phases, respectively, if we consider the time-evolution of the neutrinosphere radius. In the normal mass hierarchy, the hardness ratio of ${\nu}_{e}$ would show such a trend.

\begin{figure}[htbp]
\includegraphics[width=0.95\linewidth]{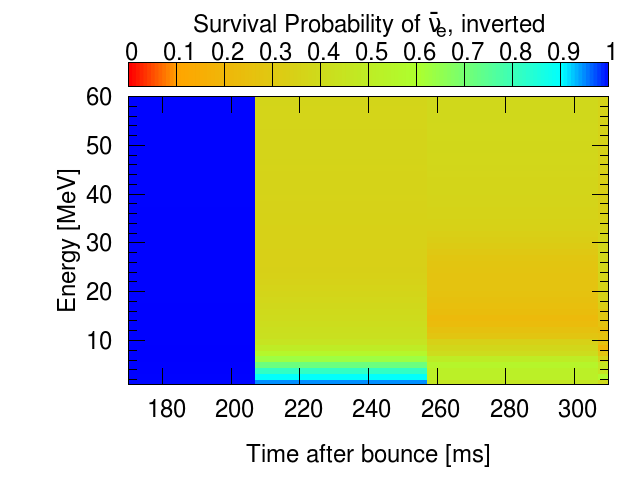}\\
\includegraphics[width=0.95\linewidth]{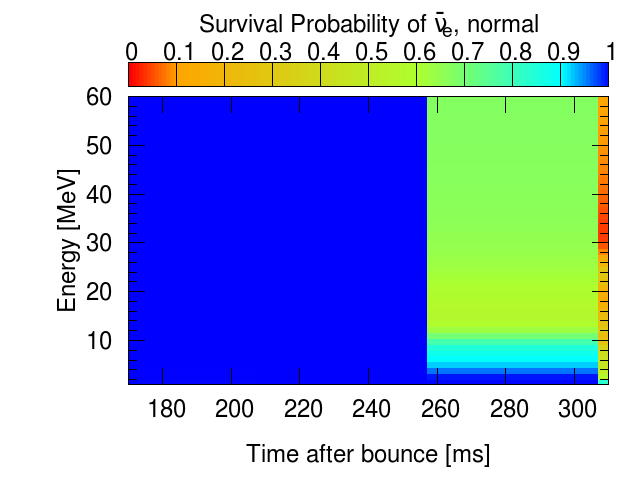}
\caption{
Same as FIG.~\ref{fig:T-E_SUV_G}, but with non-degenerate Fermi Dirac distributions for the initial neutrino spectra.
}
\label{fig:T-E_SUV_FD}
\end{figure}

\section{Fermi Dirac spectra}\label{sec:FD}
In this appendix, we show the neutrino oscillation properties using non-degenerate Fermi-Dirac (FD) spectra as the initial spectra. While we adopted a Gamma distribution \cite{Tamborra2012High-resolutionFit} as the initial spectra in the main body of this paper,  another popular choice is the FD distribution \cite{Fogli:2007bk}.  The initial neutrino spectra at the neutrinosphere is set with $\alpha_i=2$ in Eq.~\eqref{eq:gamma} and only the mean average energy of neutrinos are used to define the spectra (see middle panel of FIG.~\ref{fig:t-le} for the mean average energies).

The biggest difference from the Gamma distribution is  
the time of onset of CNO.
FIG.~\ref{fig:T-E_SUV_FD} shows the 
survival probabilities as functions of time and neutrino energy.
The top panel is for the inverted mass hierarchy.
In this case, CNO appears about 200 ms after bounce, which is 100 ms longer than the Gamma spectra (compare top panels of FIGs.~\ref{fig:T-E_SUV_G} and FIG.~\ref{fig:T-E_SUV_FD}). 

The delayed commence of CNO features with the FD spectrum also appears in the normal mass hierarchy.
The bottom panel of FIG.~\ref{fig:T-E_SUV_FD}  shows the
survival probabilities as functions of time and neutrino energy in the case for the normal mass hierarchy.
CNO features first appear at about 300 ms while the corresponding in the case of Gamma distribution is 100 ms  (see the bottom panel of FIG.~\ref{fig:T-E_SUV_G}).

\begin{figure}[htbp]
\includegraphics[width=0.95\linewidth]{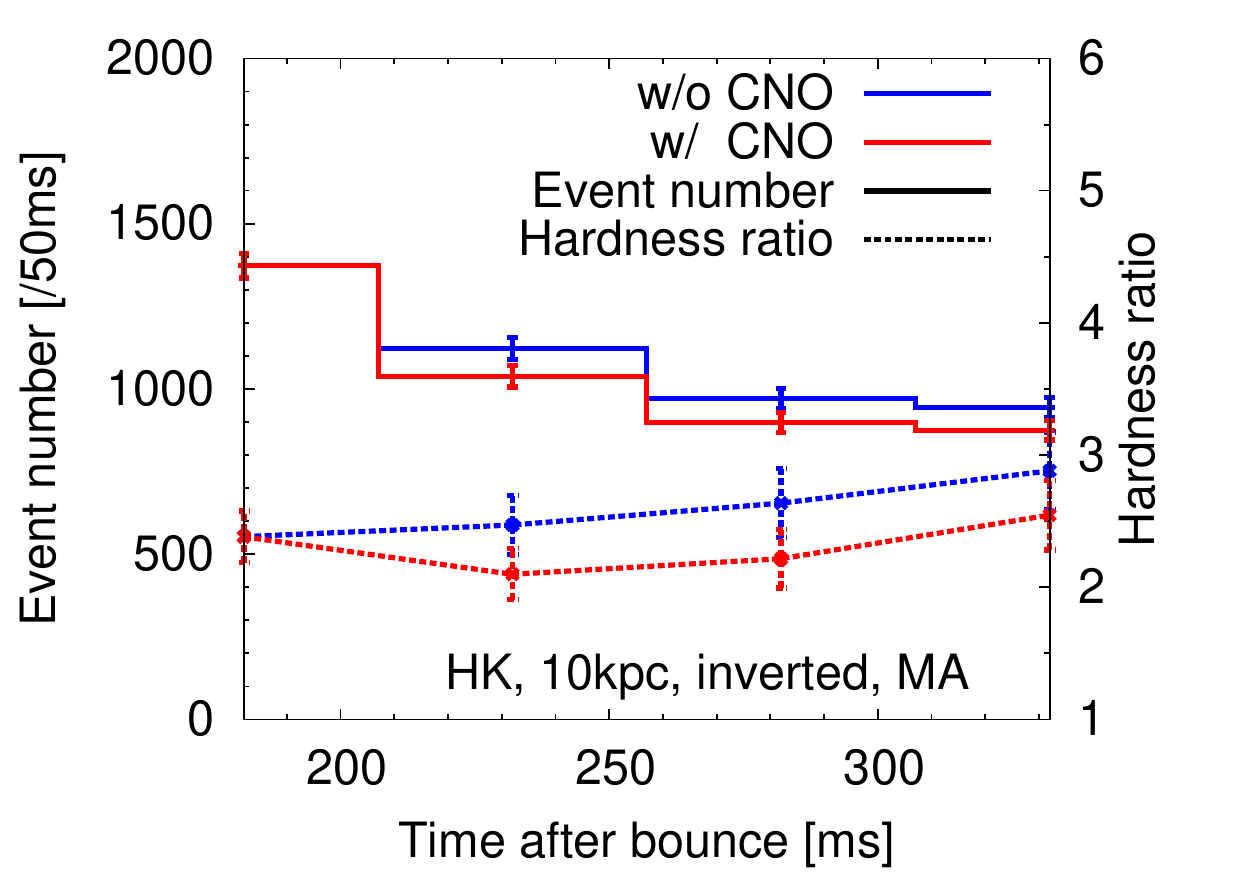}\\
\includegraphics[width=0.95\linewidth]{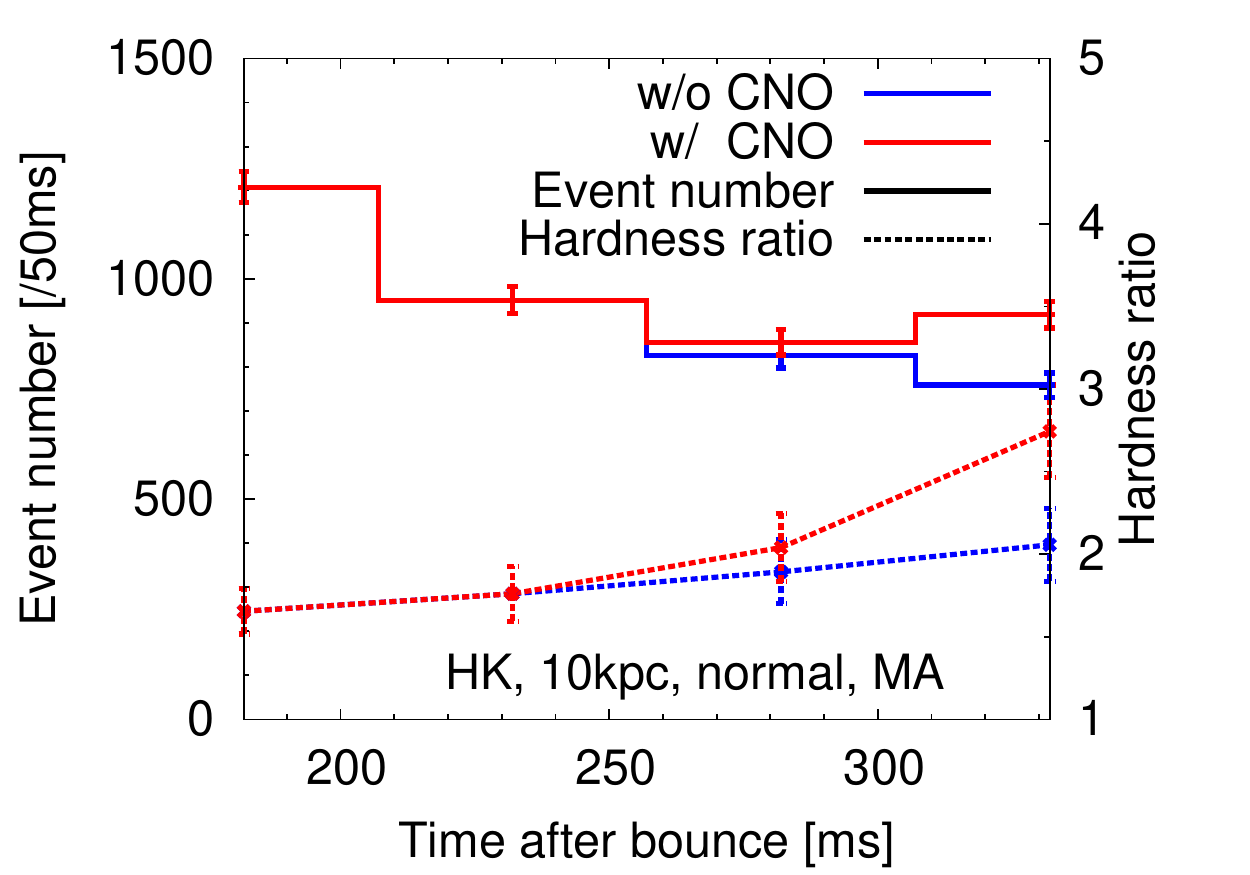}
\caption{
Same to FIG.~\ref{fig:t-N_HK_KL} but for the FD initial spectra.
Different from FIG.~\ref{fig:t-N_HK_KL}, the result for JUNO is omitted.
Note that the source distance is assumed to 10 kpc while 20 kpc is used in FIG.~\ref{fig:t-N_HK_KL}.
}
\label{fig:t-N_HK_FD}
\end{figure}

Furthermore, we evaluate the predicted event rate at HK and the hardness ratio in the case of FD spectra.
The spectrum at Earth can be obtained by Eq.~\eqref{eq:nbimh}.
The predicted event number and the hardness ratio are calculated by 
Eq.~\eqref{eq:EventRate} and Eq.~\eqref{eq:ratio}, respectively.
FIG.~\ref{fig:t-N_HK_FD} shows the time evolution of event number and the hardness ratio in 50 ms bin. 

CNO makes the spectrum of $\bar{\nu}_e$ at Earth softer in the inverted mass hierarchy.
In the top panel of FIG.~\ref{fig:t-N_HK_FD}, the hardness ratio drops after the onset of CNO.
This is qualitatively the same as for the Gamma distribution initial spectra (see top panel of FIG.~\ref{fig:t-N_HK_KL}).
Quantitatively, the impact of CNO with the FD spectra is smaller since the initial spectrum of $\bar{\nu}_e$ is more similar to that of $\bar{\nu}_X$.
The hardness ratio drops only $\sim 0.5$ in the FD case. By comparison, in the case of the Gamma distribution, the impact is more significant and the drop is $\sim 1.0$ (see top panel of FIG.~\ref{fig:t-N_HK_KL}).

On the other hand, in the case of the normal mass hierarchy 
CNO makes makes the spectrum of $\bar{\nu}_e$ at Earth harder.
The bottom panel of FIG.~\ref{fig:t-N_HK_FD} shows the evolution of the hardness ratio.
In this case, the rise of the hardness ratio is $\sim 1.0$ at 300 ms and the impact is similar to that of the Gamma distribution spectra (see middle panel of FIG.~\ref{fig:t-N_HK_KL}). 

\begin{figure}[htbp]
\includegraphics[width=0.95\linewidth]{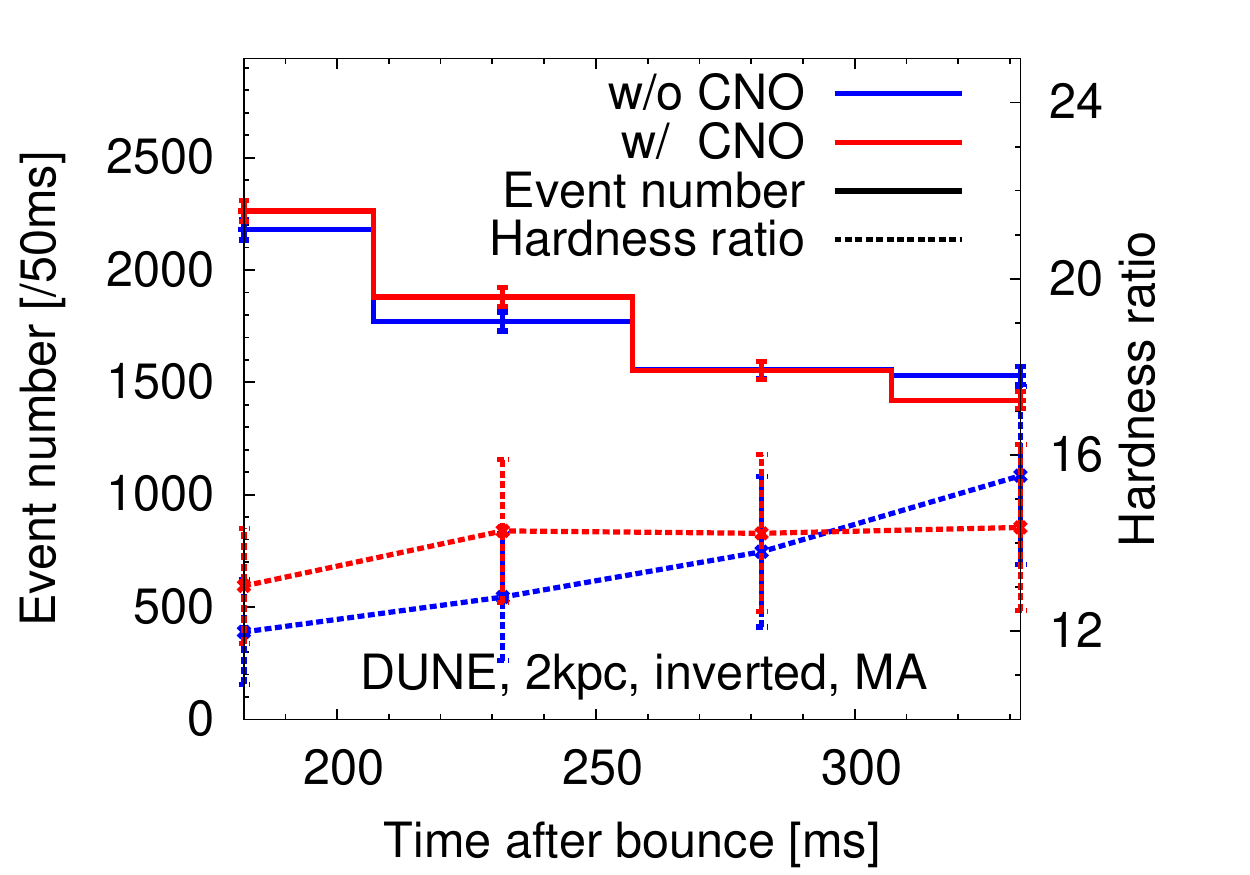}\\
\includegraphics[width=0.95\linewidth]{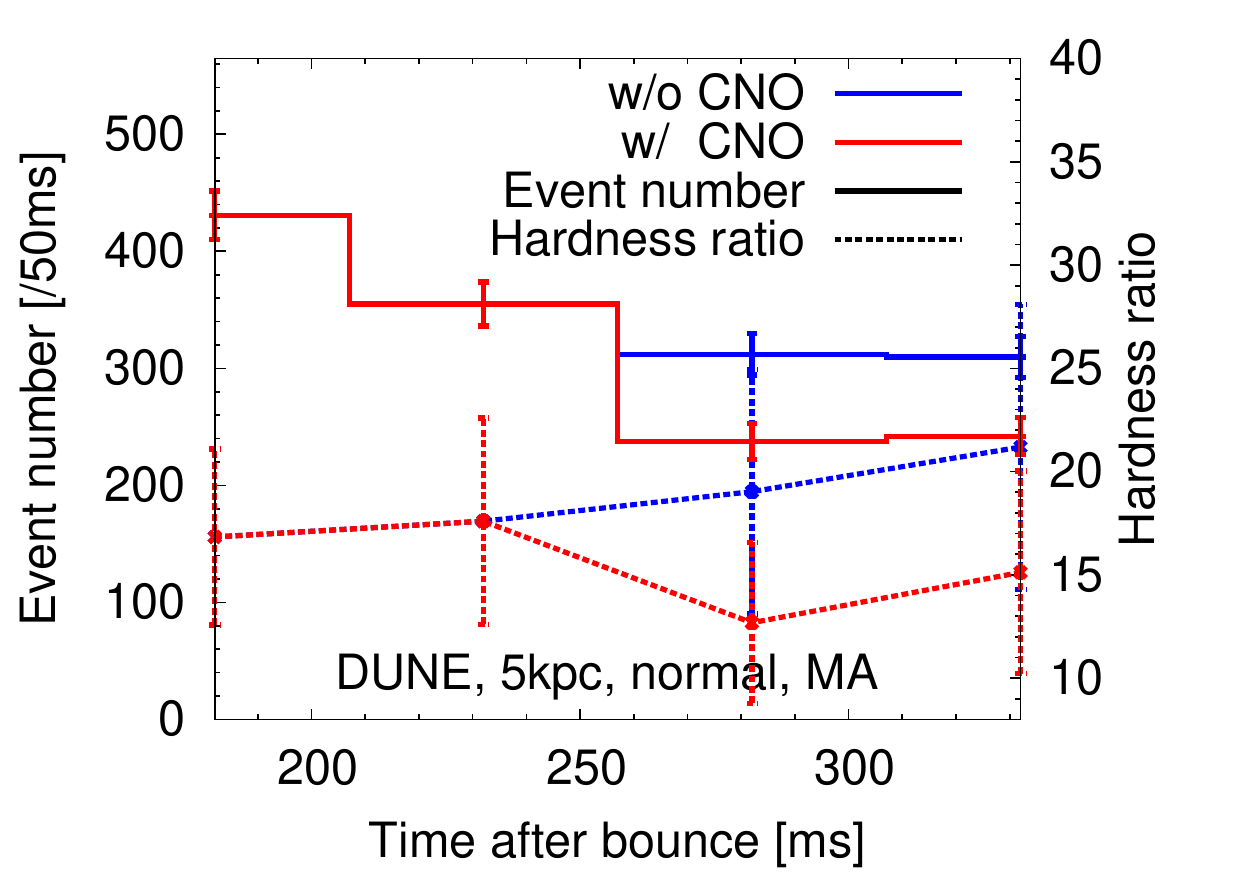}
\caption{
Same to FIG.~\ref{fig:t-NR_DUNE} but for the FD initial spectra.
}
\label{fig:t-NR_DUNE_FD}
\end{figure}

As in the main body of this article, we can also calculate CNO of $\nu_e$ using the FD initial spectra, and 
evaluate the events rate in DUNE.
The results are shown in FIG.~\ref{fig:t-NR_DUNE_FD}.
In the inverted mass hierarchy, the spectrum at Earth becomes harder when CNO starts.
That is essentially the same as the results with the Gamma distribution spectrum (see the top panel of FIG.~\ref{fig:t-NR_DUNE}). 
Similar to $\bar{\nu}_e$, the impact of CNO is smaller compared to the Gamma distribution spectrum
since the spectral shapes of $\nu_e$ and $\nu_X$ are more similar with each other when adopting the FD initial spectra.
The rise of the hardness ratio due to CNO is $\sim 2$ at 231 ms with the FD spectra and $\sim 4$ with the Gamma spectra.
The results for the normal mass hierarchy is shown in the bottom panel of FIG.~\ref{fig:t-NR_DUNE_FD}. 
Here, CNO decreases the hardness ratio of $\nu_e$. Similar to the inverted mass hierarchy, 
the impact of CNO with the FD spectra is smaller compared to that with the Gamma spectra.

\end{document}